\documentclass[fleqn,usenatbib]{mnras}

\usepackage{newtxtext,newtxmath}

\usepackage[T1]{fontenc}
\usepackage{graphicx, txfonts, natbib}
\usepackage{epsf}
\usepackage{amssymb}
\usepackage{epstopdf}
\usepackage{longtable}
\usepackage{comment}
\usepackage{subcaption}
\usepackage{float}
\usepackage{caption}
\usepackage{rotating}
\usepackage[flushleft]{threeparttable} 
\usepackage{booktabs,caption}

\DeclareRobustCommand{\VAN}[3]{#2}

\let\VANthebibliography\thebibliography
\def\thebibliography{\DeclareRobustCommand{\VAN}[3]{##3}\VANthebibliography}

\usepackage{graphicx}	
\usepackage{amsmath}	
\usepackage{amssymb}	

\newcommand{\degree}{\mbox{$^\circ$}}
\newcommand{\msun}{\mbox{$M_{\odot}$}}

\newcommand{\nick}{\mbox{$^{56}$Ni}}

\title[SN~Ia explosions and bumps in ZTF] {Constraining Type Ia supernova explosions and early flux excesses with the Zwicky Transient Factory}

 \author[Deckers et. al.]
 {M. Deckers$^1$\thanks{E-mail: deckersm@tcd.ie}, K. Maguire$^1$, M. R. Magee$^1$, G. Dimitriadis$^1$, M. Smith$^2$, A. Sainz de Murieta$^1$, A. A. Miller$^3$, 
 \newauthor
 A. Goobar$^4$, J. Nordin$^5$, M. Rigault$^6$, E. C. Bellm$^{7}$, M. Coughlin$^{8}$, R. R. Laher$^9$, D. L. Shupe$^{10}$, M. Graham$^{11}$, 
 \newauthor
  M. Kasliwal$^{11}$, R. Walters$^{11, 12}$\\
      $^1$School of Physics, Trinity College Dublin, College Green, Dublin 2, Ireland\\
      $^2$School of Physics and Astronomy, University of Southampton, Southampton, SO17 1BJ, UK\\
      $^3$Center for Interdisciplinary Exploration and Research in Astrophysics, Northwestern University, 1800 Sherman Ave, Evanston, IL 60201, USA\\
      $^4$Department of Physics, Oskar Klein Centre, Stockholm University, SE-106 91 Stockholm, Sweden\\
      $^5$ Institute of Physics, Humboldt-Universitat zu Berlin, Newton str. 15, 12489 Berlin, Germany\\
      $^6$Université de Lyon, Université Claude Bernard Lyon 1, CNRS/IN2P3, IP2I Lyon, F-69622, Villeurbanne, France\\
      $^{7}$DIRAC Institute, Department of Astronomy, University of Washington, 3910 15th Avenue NE, Seattle, WA 98195, USA\\
      $^{8}$ School of Physics and Astronomy, University of Minnesota, Minneapolis, MN 55455 \\
      $^{9}$IPAC, California Instititute of Technology, 1200 E, California Blvd, Pasadena, CA 91125, USA\\
      $^{10}$Division of Physics, Mathematics, and Astronomy, California Institute of Technology, Pasadena, CA 91125, USA\\
      $^{11}$Cahill Center for Astronomy and Astrophysics, 1200 E, California Institute of Technology, Pasadena, CA 91125, USA\\
      $^{12}$Caltech Optical Observatories, California Institute of Technology, Pasadena, CA 91125, USA\\
}

\date{Received 2021 May 05; Revised 2022 January 28; Accepted 2022 February 25}

\pubyear{2020}

\begin{document}
\label{firstpage}
\pagerange{\pageref{firstpage}--\pageref{lastpage}}
\maketitle

\begin{abstract}
In the new era of time-domain surveys Type Ia supernovae are being caught sooner after explosion, which has exposed significant variation in their early light curves. Two driving factors for early time evolution are the distribution of \nick\ in the ejecta and the presence of flux excesses of various causes. We perform an analysis of the largest young SN~Ia sample to date. We compare 115 SN~Ia light curves from the Zwicky Transient Facility to the \textsc{turtls} model grid containing light curves of Chandrasekhar-mass explosions with a range of \nick\ masses,  \nick\ distributions and explosion energies. We find that the majority of our observed light curves are well reproduced by Chandrasekhar-mass explosion models with a preference for highly extended \nick\ distributions. We identify six SNe Ia with an early-time flux excess in our \textit{gr}-band data (four `blue' and two `red' flux excesses). We find an intrinsic rate of 18$\pm$11 per cent of early flux excesses in SNe Ia at $z < 0.07$, based on three detected flux excesses out of 30 (10 per cent) observed SNe Ia with a simulated efficiency of 57 per cent. This is comparable to rates of flux excesses in the literature but also accounts for detection efficiencies. Two of these events are mostly consistent with CSM interaction, while the other four have longer lifetimes in agreement with companion interaction and \nick-clump models. We find a higher frequency of flux excesses in 91T/99aa-like events (44$\pm$13 per cent).
\end{abstract}

\begin{keywords}
Surveys-- supernovae: general
\end{keywords}


\section{Introduction}

Despite the importance of Type Ia supernovae (SNe~Ia) as cosmic distance indicators \citep{Riess1998, Perlmutter1999a, Scolnic2018}, their explosion mechanisms and progenitor systems are still debated. 
There is a general consensus that SNe~Ia are the consequence of the thermonuclear explosion of a white dwarf (WD) residing in a binary system, but the nature of the binary companion is uncertain \citep{Maoz2014, Livio2018}. There are two main proposed channels: a WD accreting mass from a non-degenerate star (SD), or two WD in a double-degenerate scenario (DD). Alternatively, theories can be categorised by the mass of the WD at the time of explosion: Chandrasekhar mass (`{$M_\textrm{Ch}$}') and sub-Chandrasekhar mass (`sub-{$M_\textrm{Ch}$}'), where the mass of the WD is less than 1.4 M$_{\odot}$ --- see \cite{Hillebrandt2013, Maoz2014, Ruiter2019, Jha2019} for comprehensive reviews. 

The specific explosion mechanism leading to SNe~Ia also remains subject to debate. The most popular model is that of the deflagration followed by the detonation of a $M_{Ch}$-WD (DDT) \cite[e.g.][]{Khokhlov1991, Khokhlov1991a}, with predictions that are generally in good agreement with observed maximum light spectra and light curves \cite[e.g][]{Seitenzahl2013}. \cite{Plewa2004} suggest gravitationally-confined detonations (GCD), where an initial off-centre deflagration ignites and expands to the surface. However, these models have a number of drawbacks, such as the overproduction of $^{56}$Ni and highly polarised ejecta \citep{Kasen2005}, conflicting with spectropolarimetric observations of SNe~Ia \citep{Wang2008, Cikota2019}. Explosions of sub-Chandrasekhar mass WDs have also been proposed. In particular, many investigations have focused on the double-detonation models \citep{Fink2007, Fink2010, Kromer2010, Sim2013, Woosley2011,Shen2014, Shen2014b, Polin2019}. In this scenario, a WD accretes a shell of He from a companion and this shell ignites and drives a shock wave into the WD, leading to the full ignition of the CO core. There are many more paths being studied that could lead to a SN~Ia such as violent mergers of two WDs \citep{Pakmor2012}, triple collision models \citep{Kushnir2013}, rotating super-Chandrasekhar mass explosions \citep{DiStefano2011}, core degenerate explosions \citep{Soker2013, Wang2017}, or the explosions of WD merger remnants \citep{Benz1990, Shen2012}. There is now significant observational evidence that there may be more than one progenitor scenario/explosion mechanism that produces `normal' SNe~Ia, e.g. studies of early light-curve variations and flux excesses \citep{Pskovskii1984, Riess1999, Conley2006, Strovink2007, Hayden2010b, Ganeshalingam2011, Firth2015, Miller2020}, ejecta masses \citep{Stritzinger2006, Scalzo2014a}, nucleosynthetic yields \citep{Seitenzahl2014, Maguire2018, Floers2019}, and the properties of SN~Ia remnants \citep{Martinez-Rodriguez2017}.

Light curves of SNe~Ia near maximum light are relatively homogeneous and do not provide the observer with sufficient information to unravel the pre-explosion conditions. However, light curves of SNe~Ia obtained within the first days of explosion enable the observer to probe the outer layers of the ejecta and the immediate surroundings of the WD, providing essential information about the accretion/merger history which are linked to progenitor scenarios and explosion mechanisms. A number of studies have investigated the early rise times of SNe~Ia (where the early evolution is parameterised as a power law with exponent, $\alpha$) to determine if it is consistent with the predictions of the fireball expansion model ($\alpha$ = 2) and found mean power law indices in the range of 1.80 to 2.40, and several SNe~Ia were clearly inconsistent with $\alpha$ = 2 \citep[][]{Conley2006, Hayden2010b, Ganeshalingam2011, Gonzalez-Gaitan2012, Firth2015, Papadogiannakis2019, Miller2020}.

The light curves of SNe~Ia are energetically driven by the radioactive decay of $^{56}$Ni, which is synthesised in the explosion \citep{Arnett1982, Pinto2000, Piro2013}. The $^{56}$Ni distribution has significant effects on the dark phase \citep[the time that passes between the explosion epoch and time of first light;][]{Piro2013} and the shape of the early rise \citep{Gamezo2005, Piro2013, Mazzali2014, Piro2016, Magee2018}. If $^{56}$Ni extends to the outer regions of the ejecta, a shallower, earlier rise, and a shorter dark phase is expected. The full range of rise indices can be reproduced by simply varying the $^{56}$Ni distributions \citep{Piro2013, Magee2018, Magee2020}. \cite{Dessart2014} argued that a pulsation delayed detonation of a WD can also produce light curves that are brighter and bluer at earlier times, synonymous to the effects of a highly mixed ejecta. However, early time spectra and knowledge of the velocity gradients can be used to distinguish between these two scenarios \citep{Dessart2014}. 

In a handful of cases, a flux excess in the early light curves of SNe~Ia has been detected (commonly referred to as `bump' or `flash') e.g.~SN 2012cg \citep{Marion2016}, SN 2014J \citep{Goobar2015}, iPTF14atg \citep{Cao2015}, SN 2016jhr \citep{Jiang2017}, SN 2017cbv \citep{Hosseinzadeh2017}, SN 2018oh \citep[][]{Dimitriadis2019, Shappee2019, Li2019}, SN 2019yvq \citep{Miller2020b}, SN 2020hvf \citep{Jiang2021}, and SN 2021hpr (Lim et al., in prep.). \cite{Olling2015} analysed a sample of three SNe~Ia with extremely high cadence Kepler light curves, and found no evidence of flux excesses at early times in these events. \cite{Magee2020} (hereafter M20) investigated a literature sample of SNe~Ia with early light curve data and found that 22~per cent of objects required additional flux at early times to fit the light curve. \cite{Jiang2018} coined the term `Early-broad EExSNe~Ia', including objects with very broad early light curves in the class of excess flux objects (EExSNe). Under this broader definition, \cite{Jiang2018} found that all 91T/99aa-like events could be classified as EExSNe.  

A number of different scenarios have been suggested to produce these flux excesses, such as interaction with a non-degenerate companion star \citep{Kasen2010}, interaction with circumstellar material \citep[CSM;][]{Piro2016, Kromer2016a}, signatures of a He-shell detonation \citep{Noebauer2017,Jiang2017,Polin2019, Magee2021}, or the presence of \nick\ clumps in the outer ejecta \citep{Dimitriadis2019, Shappee2019, Magee2020a}. However, linking an observed flux excess to one particular scenario has proven difficult. The main reason for this is that while some of these models are able to reproduce the light curves with flux excesses, the accompanying model spectra are usually not in agreement with observations. Both the early and maximum light spectra can be significantly affected by He-shell detonations or \nick\ clumps \citep[e.g.][]{Maeda2018a, Polin2019, Magee2020a, Miller2020b}, and signatures expected from the interaction scenarios (e.g. H and He in nebular spectra) have only been observed in a very limited number of cases \citep{Kollmeier2019, Prieto2020}.

Our aim in this work is to extend the study of M20 by analysing the $^{56}$Ni distributions of a large, homogeneous sample of SNe~Ia from the Zwicky Transient Facility \citep[ZTF;][]{Bellm2018, Graham2019, Masci2019, Dekany2020} to determine if the Chandrasekhar-mass models of M20, with varying \nick\ masses, \nick\ distributions, shapes of the density profile, and kinetic energies, can reproduce the observed light curves. We also aim to constrain the fraction of objects in this sample that display a flux excess. The general properties, rise times and colour evolution of the sample of SNe~Ia in this sample are described in \cite{Yao2019, Miller2020} and \cite{Bulla2020}. In Section \ref{sampleselection}, we describe the ZTF SN~Ia sample, as well as the model grid of M20. In Section \ref{analysis}, we discuss our methods for comparing the observed light curves to the model grid and identifying flux excess in the early light curves, as well as an efficiency analysis of the flux excess detection. In Section \ref{results}, we present the results of our light curve analysis, and in Section \ref{discussion}, we discuss the implications of our results in terms of explosion mechanisms and the origin of flux excesses. We conclude our work in Section~\ref{conclusion}. 

\section{Sample Selection and Model Grid}\label{sampleselection}

Our aim in this work is to compare the 2018 ZTF SN~Ia sample to a grid of 300 Chandrasekhar-mass models presented in M20. The models are parameterised with different $^{56}$Ni masses and $^{56}$Ni distributions, kinetic energies, and density profiles. In Section~\ref{dataprocessing}, we introduce the sample of SN~Ia light curves used in this analysis, and discuss how we processed the light curves in order to compare them to the model light curves. In Section~\ref{models}, we present a description of the M20 model grid and in Section~\ref{cuts} we describe the cuts we applied to ensure our sample has sufficiently early-time data and consists solely of `normal' SNe~Ia \citep[those that follow the width-luminosity relation;][and can be used in cosmological analysis]{Pskovskii1977, Pskovskii1984, Phillips1993}. Finally, in Section~\ref{galaxies} we discuss our identification of the host galaxies for the sample and their general properties.

\begin{figure}
    \centering
    \includegraphics[width=8cm]{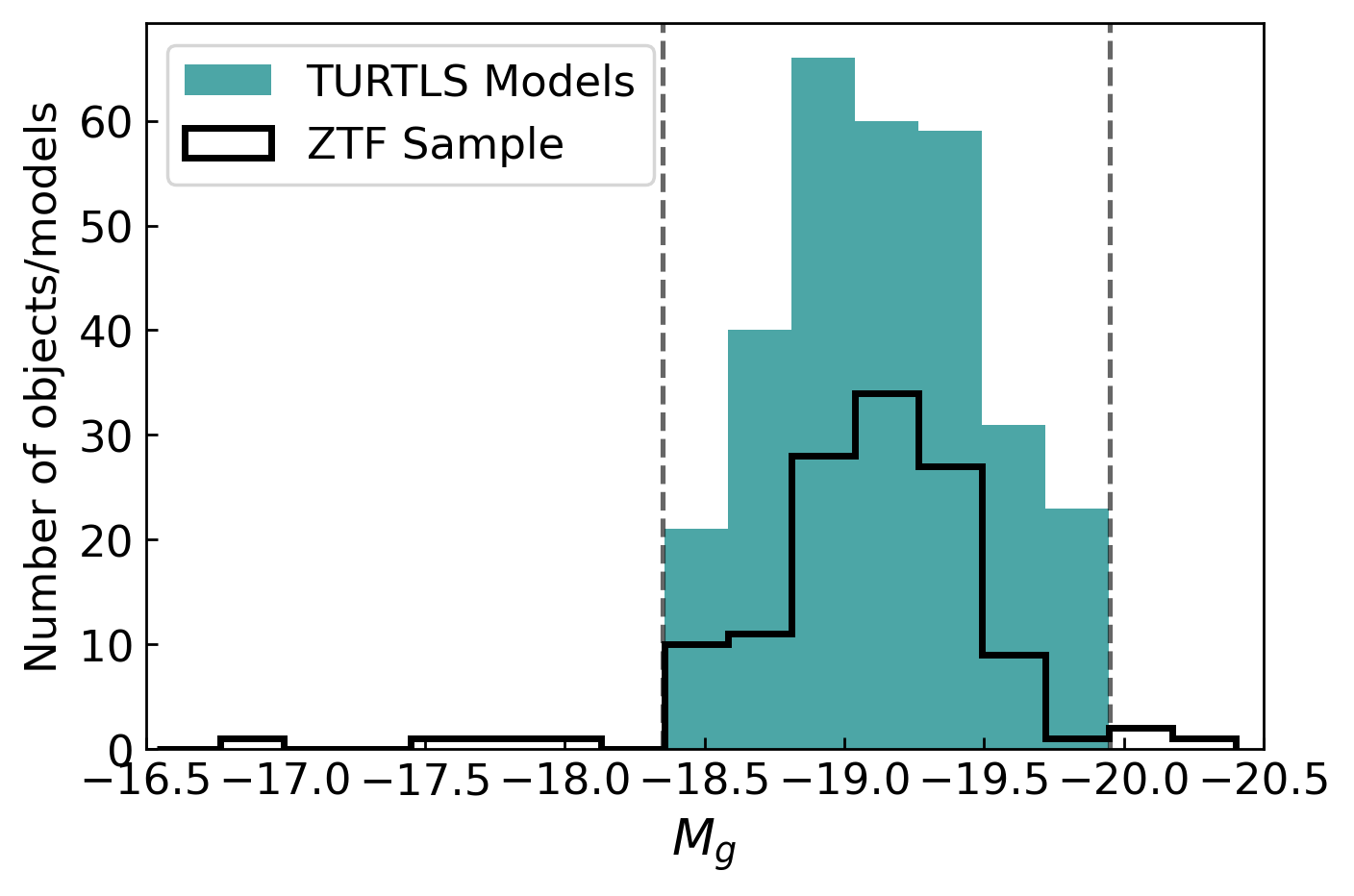}
    \caption{A plot showing the distribution of \textit{g}-band magnitudes at maximum light covered by the \textsc{turtls} model grid, and those derived from SALT2 fits by \protect\cite{Yao2019} for the observed 2018 ZTF SN~Ia sample before making any of the cuts described in Section \ref{cuts}. The dashed lines indicate the absolute magnitude cuts we applied at -19.95 $\leq$ peak \textit{g} $\leq -18.35$.}
    \label{g_band_distribution}
\end{figure}

\subsection{Data Processing of ZTF light curves}\label{dataprocessing}

The ZTF collaboration conducted a high-cadence extra-galactic survey during its first year which obtained six epochs (3\textit{g} + 3\textit{r}) nightly over a sky coverage of 2500 deg$^2$ \citep{Bellm2019, Dekany2020}, resulting in 336 spectroscopically classified, high cadence SNe~Ia light curves \citep{Graham2019}\footnote{All public data is available on the GROWTH Marshall \citep[][\url{http://skipper.caltech.edu:8080/cgi-bin/growth/marshal.cgi}]{Kasliwal2019}}. The high-cadence and multi-band photometry allows us to investigate the colour evolution, as well as the luminosity evolution of these objects from a very early stage. 
\cite{Yao2019} describe the general properties of a sub-sample of the 2018 ZTF data set, which consists of 127 SNe~Ia with both \textit{g}- and \textit{r}-band observations at least 10~d before peak \citep[obtained originally through ztfquery;][]{Rigault2018}. All objects have a minimum of five observations in both bands before the time of peak in the \textit{B}-band.

In order to compare the ZTF light curves to the model grid of M20, we need to convert the raw counts to absolute flux. Following the prescription presented in section 3.5 in \cite{Yao2019}, we convert the forced photometry counts to flux values for all 127 objects using the photometric zero point of every image provided by the pipeline, and correct the baseline by the offset values calculated in \cite{Yao2019}.
We apply an error floor of 2~per cent to the data to account for underestimated photometric uncertainties. The flux on each specific epoch has been stacked, with means weighted proportional to the size of the uncertainties. We also correct for time dilation, placing the light curves in the SN rest frame (for the rest of this paper, all quoted times have been corrected to the SN rest frame).

We apply the Milky Way extinction correction in the direction of each SN using dust maps from \cite{Schlafly2011}, assuming $R_V$ = 3.1 and using the CCM extinction law \citep{Cardelli1989}. \cite{Bulla2020} performed SNooPy fitting \citep{Burns2014} and derived host galaxy reddening values and K-corrections for 65 objects in the sample. The host extinction values derived by \cite{Bulla2020} range from 0.008--0.78 mag, with 10 objects having a host extinction greater than 0.3 mag. Following \cite{Yao2019} and \cite{Miller2020}, we do not correct for host extinction, and any highly reddened objects are excluded based on the colour ($c$) derived from fitting the light curves with the light-curve fitter SALT2 \citep[See Section \ref{cuts};][]{Guy2007} since SALT2 does not explicitly fit for host reddening, but it is encapsulated in the $c$ value. K-corrections at early times for SNe~Ia are uncertain as they are derived by extrapolating from 15~d prior to maximum light to earlier epochs. Moreover, \cite{Bulla2020} found that the K-corrections for their subset of the ZTF SN~Ia sample were small, with the objects at the highest redshifts having K-corrections of $\leq 0.1$ mag. Therefore, no K-corrections have been applied in the following analysis.

\subsection{Models}\label{models}

For a detailed description of the models used in this work, we direct the reader to \cite{Magee2018}, and M20. Briefly, the model grid contains 300 light curves which were produced using the one dimensional, Monte Carlo radiative transfer code, \textsc{turtls} \citep[][M20]{Magee2018}. The code is designed to model the light curves of thermonuclear SNe from explosion up to maximum light. Our particular suite of models was produced by varying the ejected $^{56}$Ni mass, the $^{56}$Ni distribution, the shape of the density profile, and the kinetic energy of the ejecta.  We implement the Chandrasekhar-mass model suite of M20 of 255 models with $^{56}$Ni masses of 0.4, 0.6 and 0.8 \msun, which was used to fit the light curves of 35 well-observed SNe~Ia in M20. We supplement this model suite with 45 additional models with a Ni mass of 0.5 calculated as part of this work, in order to increase the resolution in the 0.4--0.6 $M_{\odot}$ region because our initial analysis showed most objects to be best fit by models in this range.

The shape of the $^{56}$Ni distribution as a function of the mass coordinate, $m$, is defined using the relation of \cite{Magee2018},
\begin{equation}
^{56}\textrm{Ni} \left(m\right) \textrm{=} \frac{1}{\exp \left(P\left[m-M_{Ni}\right] \textrm{/}M_{\odot}\right) \textrm{+} 1} ,
\end{equation}\label{equation}

where $M_{Ni}$ is the total $^{56}$Ni mass in $M_{\odot}$, and \textit{P} parameterises the shape of the ejecta  \citep[defined as \textit{s} in][]{Magee2018}, with smaller values of \textit{P} representing more extended $^{56}$Ni distributions and larger values of \textit{P} representing more compact distributions ($^{56}$Ni confined to the core). The model grid contains two density distributions: exponential (EXP) or double power law (DPL). There are four ejected $^{56}$Ni masses (0.4, 0.5, 0.6, 0.8 $M_{\odot}$), nine kinetic energies ($0.50$ -- $2.18 \times 10^{51}$ erg), and five \textit{P} values (spanning $3-100$ in log space). The model names encapsulate all the information about the parameters (e.g. EXP\_Ni0.4\_KE0.65\_p4.4).

It is important to note that the model composition is simplified to three ejecta zones: the innermost zone is composed of iron group elements (IGEs, assumed to be 100~per cent $^{56}$Ni at the time of explosion), a middle zone of intermediate mass elements (IMEs), and an outer layer composed of $ 0.1\ M_{\odot}$ of carbon and oxygen. The assumption of a pure radioactive $^{56}$Ni composition in the inner zone at the time of explosion results in an overestimation of the brightness compared to a more physical composition. The $^{56}$Ni masses derived should therefore be considered as lower limits. Because the models assume local thermal equilibrium (LTE), they only predict the light curve evolution accurately up to approximately maximum light. Correspondingly, we only use light curve data ranging from 0--20~d post explosion for our fitting.

\subsection{Sample cuts}\label{cuts}

\begin{table}
  \begin{center}
    \caption{Summary of the cuts applied to the observed SN~Ia sample.}\label{tablecuts}
    \begin{tabular}{l|c}
    \hline
      Condition & No. of SNe~Ia remaining \\
      \hline
      \hline
      Starting Sample & 127 \\
      $-0.3 \leq \  c \ \leq 0.3$  & 122 \\
      $-3.0 \leq \ x_1 \  \leq 3.0$ & 119 \\
      $-19.95 \leq$ peak \textit{g} $\leq -18.35$ & 116 \\
      First detection $\leq$ peak (g) $-$ 14~d  & 115\\
    \hline
    \end{tabular}
  \end{center}
\end{table}

This paper focuses on `normal' SNe~Ia to match the parameter space covered by the M20 model grid, which does not include the extremes of the SN~Ia brightness distribution. Accordingly, we apply a number of cuts to our initial sample of 127 SNe~Ia to remove peculiar objects. We have adopted the SALT2 parameters determined by \cite{Yao2019} and applied the cosmological cuts as quoted in \cite{Smith2020} ($-0.3 \leq  c  \leq 0.3$, $ -3.0 \leq  x_1  \leq 3.0$), reducing the sample down to 119 objects. These cuts removed one Ia-CSM (a SN~Ia interacting with CSM, ZTF18aaykjei), three over-luminous `super-Chandrasekhar-mass'/03fg-like objects (`SC', ZTF18abdpvnd, ZTF18aawpcel, ZTF18abhpgje), one 91T-like (ZTF18abealop), and three normal objects (ZTF18aasesgl, ZTF18abwdcdv, ZTF18abgmcmv), with spectroscopic classifications adopted from \cite{Yao2019}. 

For a fair comparison between data and models we only include SNe~Ia with absolute peak magnitudes covered by the M20 model grid, so any objects with an absolute peak \textit{g}-band magnitude brighter than $-$19.95 mag or fainter than $-18.35$ mag are removed (see Fig.~\ref{g_band_distribution}). This cut further excludes one 02cx-like (ZTF18abclfee), one normal (ZTF18aansqun), and one 99aa-like (ZTF18abjdjge) from our sample reducing the number to 116 SNe~Ia. 

M20 analysed the effects of limiting early-time data and concluded that in order to reliably constrain the $^{56}$Ni distribution, a first 3-$\sigma$ detection within $\sim$3\,d of the explosion epoch (approximately equivalent to 14~d prior to \textit{B}-band peak) is required. To satisfy this requirement, we check that the SNe~Ia in our sample have a first 3-$\sigma$ detection in the \textit{g}- or $r$-band at least 14~d before the time of the estimated \textit{g}-band maximum derived from SALT2 fitting (in the SN rest frame). After applying these cuts, we are left with 115 SNe~Ia in our final sample for investigation (see breakdown of cuts in Table~\ref{tablecuts}). Our final sample still includes a number of more unusual SN~Ia classes: nine 99aa-like SNe, two 91T-like SN, one 86G-like SN, and one `SC'/03fg-like SN. 

\subsection{Host galaxies}\label{galaxies}

For each SN in our sample, the host galaxy was identified from PanSTARRS legacy imaging \citep{Chambers2016} using the 'Directional Light Radius' method \citep[DLR;][]{Sullivan2006, Gupta2016}. In detail, the closest galaxy (in units of DLR) of the SN position is considered the host galaxy of each event with a requirement of DLR<10 \citep{Gupta2016}. SN with no host meeting this criterion (6 SNe Ia, 5.2~per cent) are considered hostless. 
For each host galaxy, \textit{griz}-band fluxes were determined from the PanSTARRS images using the \texttt{SEP} source extraction code \citep{Barbary2018}, based on Source Extractor \citep{Bertin1996}. 
To estimate the stellar mass and star-formation rate of each host we use the methodology described in \citet{Sullivan2010} and \citet{Smith2020}. We use the \texttt{P\'EGASE.2} spectral synthesis code \citep{Fioc1997} combined with a \citet{Kroupa2001} initial mass function (IMF) to calculate the spectral energy distribution (SED) of a galaxy as a function of time using 9 exponentially declining star-formation histories (SFH) combined with 7 foreground dust screens. To break the degeneracy between age and metallicity, for each SFH we generate 4000 models, each with a burst of star-formation randomly superimposed between 1 and 10 Gyr after formation \citep{Childress2013}. Each model is then evaluated at 102 time steps. 
For each host, we compare our \textit{griz} fluxes, after correcting for Milky Way extinction, to those of each model SED at the redshift of the SN to find the best fitting template and stellar mass. Uncertainties on stellar mass are calculated using a Monte-Carlo approach. The mean host galaxy stellar mass for our sample is 9.82 M$_\odot$, and host masses and DLR values are given in Table \ref{galaxy_data}.

\section{Methods}\label{analysis}

In this section, we discuss the fitting routine implemented to compare the model light curves to our data set of observed SN~Ia light curves with early \textit{g}- and \textit{r}-band data. We describe the conditions for classifying objects as `well fit', and compare the data to objects presented by M20 to ensure the fitting routine is not affected by differences between the samples. The method for detecting flux excesses is described in Section \ref{bump_detection}, and the efficiency of this method is investigated in Section \ref{efficiency}.

\subsection{Comparing observed light curves to the model grid}\label{comp_description}

\begin{figure}
    \centering
    \includegraphics[width=8.cm]{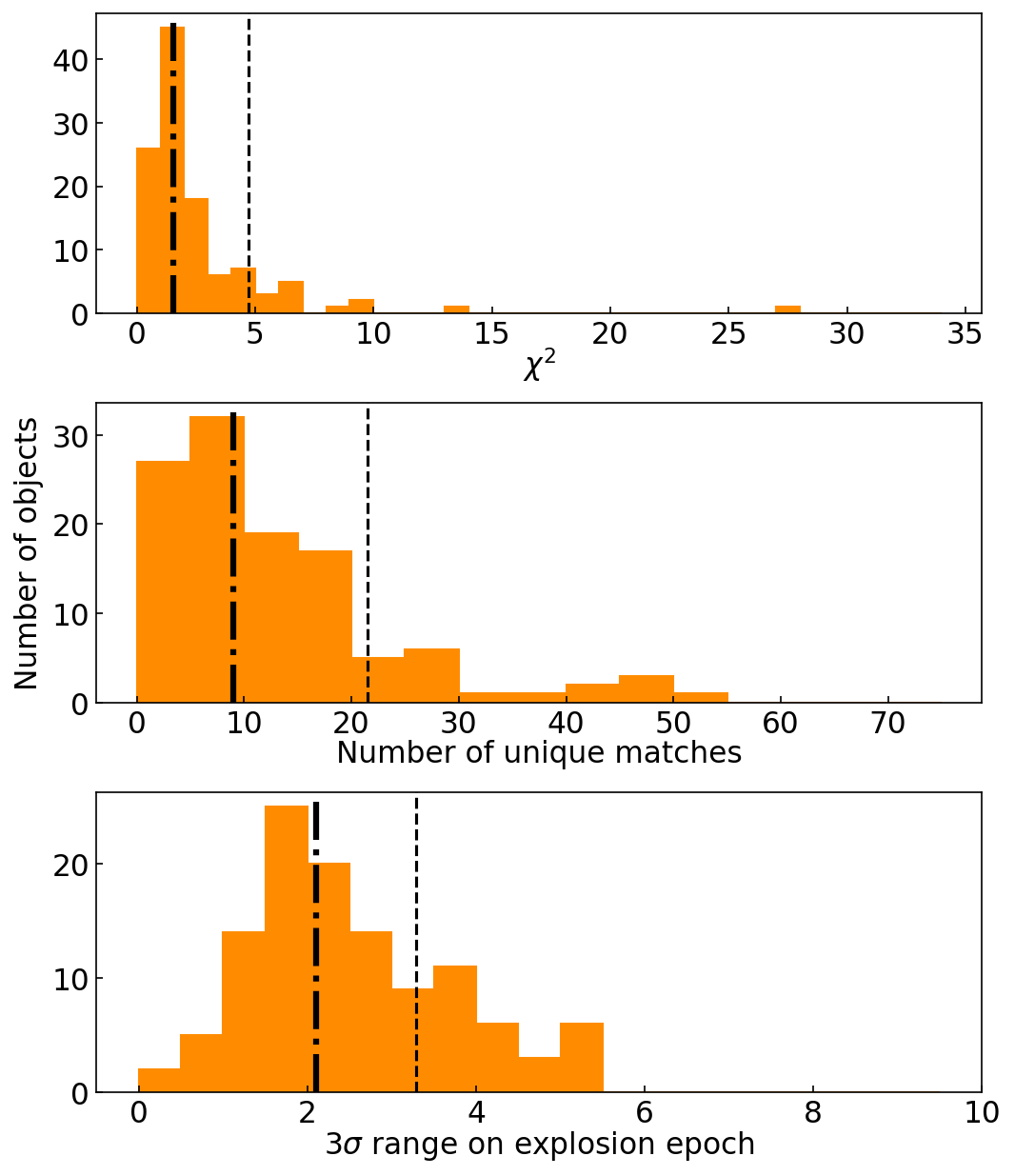}
    \caption{Histograms showing the distribution of $\chi^2$ values, the number of unique fits, and the 3-$\sigma$ range for the estimated explosion dates for each object. The vertical dot-dashed lines indicate the median of the distribution, and the dashed lines indicate one standard deviation from the median where we have placed the cut to separate well from badly fit objects.}
    \label{cuts_histo}
\end{figure}

We compare each observed SN~Ia light curve to the grid of Chandrasekhar-mass models of M20 in flux space using the $\chi^2$ metric. Performing the fits in flux space has the benefit that we can include all the available data, including lower significance early data points. Following the methodology from M20, the first `detection' is taken as the earliest $3 \sigma$ detection in the \textit{g}- or \textit{r}-band. A rough date of maximum light is initially estimated as the mean of the epochs of peak magnitude in the \textit{g}- and \textit{r}-bands. We limit the range of data for fitting from 30~d before maximum light up to 20~d after first detection. We then fit simultaneously for the distance modulus, explosion date, and the best-fitting model. 

We allow the distance modulus to vary up to 0.3 mag to account for the coarseness of the model grid, as well as potentially underestimated distance uncertainties. We iterate over the distance modulus range in 15 steps. As described in M20, it is assumed that the explosion could have occurred anytime between 30~d before maximum light, up to the time of the first 3-$\sigma$ detection. The code iterates over this range in increments of 0.1 d.
The $\chi^2$ value is computed for each comparison of the model with the SN~Ia light curve, for all distance modulus, and explosion date in the allowed ranges. Subsequently, we calculate $\chi^2_{red}$ by dividing $\chi^2$ by the degrees of freedom (DOF), estimated as the number of data points in each fit, minus the number of parameters fitted. We normalise the $\chi^2_{red}$ values by dividing by the smallest $\chi^2_{red}$ (such that the best-fitting model has a $\chi_{norm}^2$ = 1.0). Fits with $\chi^2_{norm}$ values that fall within the 3-$\sigma$ confidence region of $\chi^2_{norm}$ = 1, where the cut off value is taken from the percent point inverse function dependent on the DOF, are considered as potential matches. The 3-$\sigma$ range is used as a measure for the uncertainty of the model parameters, distance moduli, and explosion epochs.

M20 consider an object consistent with the model if the mean residual is approximately zero, and the maximum value of the residual in each band is $\leq$ 1 mag. We base our well fit criteria on the distribution of the $\chi^2$ value, the number of unique fits within the 3-$\sigma$ range (the number of unique matches corresponds to the number of models from the grid which are matched to an object), and the 3-$\sigma$ range on the explosion date. We find the median of each distribution (shown as the dash-dot line in Fig.~\ref{cuts_histo}), and consider the upper limit as one standard deviation from the median (shown as the dashed line in Fig.~\ref{cuts_histo}). It should be noted that this study does not aim to constrain the intrinsic rate of $M_{Ch}$-mass explosions, but rather it aims to test whether variation in the \nick\ distribution is able to account for significant variation in the early light curves of normal SNe~Ia. Therefore, the purpose of these cuts is to remove the tails of the distribution, leaving us with a normal distribution in these three parameters to describe the well fit sample where these values vary randomly due to the uncertainties in the data. The limits are: a $\chi^2$ <= 4.7, <= 22 unique potential matches, and a 3-$\sigma$ explosion date range <= 3.3 d. The latter two conditions relate to the uniqueness of a fit, and how well the model parameters are constrained. Noise in the data means we are unable to constrain the model, but we are not able to rule out a Chandrasekhar mass explosion with certainty. These parameters are also highly dependent on the resolution of the model grid, but since we are comparing fits to the same model grid, the relative values are still useful. When using this method to compare light curves to an expanded model grid, these limits should be adapted. Fits with high $\chi^2$ values indicate that we are unable to fit the data with our models. Combined, these quality cuts remove 37 SNe~Ia from the sample. This does not necessarily imply that all these SNe Ia originate from non-Chandrasekhar mass explosions, but rather that we are unable to constrain model parameters from the fit. This could be due to poor data quality, or intrinsic differences in the light curves compared to the models. See Section \ref{badfittext} for further discussion of these events. Our final `well-fit' sample of 78 SNe~Ia has a mean and standard deviation $\chi^2_{red}$ = $1.6 \pm 1.0$, $8 \pm 5$ for the number of unique fits, and $1.9 \pm 0.7$ d for the 3-$\sigma$ range of explosion dates.

\subsection{Validation of the use of ZTF \textit{gr}-band light curves}\label{tests}

We analyse the potential effects of the differences between our sample and that presented by M20. We investigated whether the availability of only two bands (\textit{g} and \textit{r}) in the ZTF sample may affect our results, compared to those with multi-band coverage in the M20 sample. We do this by selecting objects in M20 where the earliest data is in the \textit{g}- or \textit{r}-band but also have wider wavelength coverage, finding 7 objects. We then refit these objects, limiting the bands to just \textit{g}- and \textit{r}- bands (Sloan \textit{g}- and \textit{r}-bands, which are comparable to ZTF bands). For five events (SN 2011fe, SN 2013gy, iPTF16abc, SN 2017cbv, SN 2017erp), the best-fitting model was not affected by reducing the number of bands available for fitting. Two events (iPTF13ebh, iPTF13dge) change from an original best fit of  $P$ = 9.7 to $P$ = 4.4 when using only the \textit{g}- and \textit{r}-band data. However, for iPTF13ebh only the best match changes, but the next best match is still included in the 3-$\sigma$ range. In the case of iPTF13dge, we find that it has a large gap (4~d) between predicted explosion date and the epoch of the first data point. In this case it is not possible to narrow down the explosion range sufficiently, and subsequently the model parameters are badly constrained. Not all the light curves in our sample have a 3-$\sigma$ detection prior to 3~d after explosion, but all have at least one $\leq$ 3-$\sigma$ detection prior to this epoch which we also include in the fitting. If there is insufficient early data to constrain the explosion epoch, the fit will not pass our quality cut on the 3-$\sigma$ range of explosion dates. We conclude that \textit{g}- and \textit{r} band data as in the ZTF sample are sufficient for constraining the \nick\ distribution of SNe~Ia when using the \textsc{turtls} models.

\subsection{Detection of flux excess} \label{bump_detection}

SNe~Ia with flux excesses in their early light curves are commonly presented in single object studies. In these cases, high-cadence observations at early times allow the bump to be clearly resolved \citep[e.g. SN 2017cbv;][]{Hosseinzadeh2017} or a data point is obtained during the decline of the bump \citep[e.g. SN 2019yvq;][]{Miller2020b}, separating the rise of the bump from the $^{56}$Ni-powered rise to the secondary peak. \cite{Jiang2018} were the first to systematically look for early flux excesses in SN~Ia light curves. In their study, light curves with prominent flux excesses are identified by eye or their presence is taken from the literature, and early-broad EExSNe are identified through a comparison with the light curve of SN 2012cg. M20 search for flux excesses in the early light curves of SNe~Ia in a more quantitative way by requiring that a data point produces a residual $> 1$ mag when compared to the best matched model. 
We aim to search for flux excesses in a large sample of SNe~Ia and want to use a quantitative reproducible method. To this end, we implement a similar method as presented in M20. However, we assess the residuals in flux rather than magnitude space so that all data points including upper limits can be analysed in a consistent fashion. 

We base our parameterisation of potential flux excesses in the data on the theoretical predictions from progenitor models. The additional power sources that are predicted to produce early flux excesses are expected to be blue \citep[e.g.][]{Kasen2010,Polin2019}. However, this additional luminosity must be combined with the underlying Ni-powered light curve that can make the early colour appear less blue \citep[e.g.][]{Piro2016}. Therefore, we check for potential flux excesses in both the \textit{g}- and \textit{r}-bands. The longest flux excess in the \textit{B}-band in the interaction models of \cite{Kasen2010} lasts for $\sim$6 d and we search for flux excesses of up to this length.  Specifically, we search for a positive residuals between the data and best fit model in the six days after the explosion epoch estimated from the best model fit, with an uncertainty on the explosion epoch corresponding to the 3-$\sigma$ model range. This range also covers the timescales of flux excesses seen in previous literature events \citep[e.g.][]{Hosseinzadeh2017, Dimitriadis2019, Shappee2019, Miller2020b}. We require two consecutive epochs to have non-zero residuals in a single band in this range, where at least one has a residual of $>$2 per cent of the peak flux\footnote{ This choice is motivated by the strength of early flux excesses seen in literature but the exact value is somewhat arbitrary. However, we consider it a conservative value that only selects the highest significance events. This method assumes that the brightness of a flux excess should scale with the overall brightness of the SN, which is not always the case, since the brightness of the flux excess in the case of companion interaction depends on external properties (e.g.~companion separation, viewing angle) that are not inherently related to the exploding WD.} and also require all models in the 3-$\sigma$ range to have a non-zero residual outside their uncertainties at these epochs. Models of companion interaction \citep{Kasen2010}, CSM interaction \citep{Piro2016}, and \nick -clumps \citep{Magee2020a} all predict flux excesses that begin at approximately the same time as the explosion. The estimated explosion epoch from our fit is highly uncertain when a flux excess is present, but we rule out flux excesses that begin later than 1 d after the maximum explosion epoch in the 3-$\sigma$ range. Some ZTF light curves show unphysical scatter that is most prominent around peak brightness so again to be conservative in what we consider an early flux excess, we remove any objects with potential flux excesses at early times that also show a scatter of similar significance at later times (any days $>$10 d after explosion). Because this may remove events with true early flux excesses, for consistency we remove any SNe~Ia in the full sample with similar levels of scatter prior to performing the flux excess rate calculation in Section \ref{excess_rates}.

We test the sample for contamination from false excesses through a search for negative flux excesses, performed by inverting the flux excess criteria. No negative flux excesses are found, indicating that the method can robustly detect true excesses.

We also considered an alternative method of removing the data at early epochs, refitting the light curve, and checking if the early data produces a positive residual relative to the best match. However, M20 emphasise the importance of the early data for reliably constraining the \nick\ distribution, and we clearly notice the effect of the missing data. By leaving out the early epochs, we identify 70~per cent of objects as having a flux excess, but upon closer analysis we find that the fits show a preference for models with more compact \nick\ distributions. Consequently, the data points corresponding to the dark phase of the models are picked up as excess flux. Both methods have drawbacks, but we opt for the method described in detail above, as this method is less likely to produce false positives. 
It should be noted that the explosion parameters derived from a model fit, as well as the estimated explosion epoch, are not well constrained in the scenario where a flux excess is detected and these should not be used for further analysis.

\section{Results}\label{results}

Following the methodology presented in Section~\ref{comp_description}, we present the results of the analysis here. We first describe the objects that passed the quality cuts and appear to be well fit by Chandrasekhar-mass explosion models, and present the distributions of the model parameters in Section~\ref{wellfittext}. This is followed by an analysis of objects that have no good match in our model grid in Section~\ref{badfittext}. Lastly, in Section~\ref{excess} we present our sample of SNe~Ia with a potential flux excess, their light curves, host galaxy properties, and the details and results of an detection efficiency calculation.

\subsection{SNe~Ia well fit by Chandrasekhar-mass explosions} \label{wellfittext}

\begin{figure}
\centering
\begin{subfigure}{9cm}
\centering\includegraphics[width=9.cm]{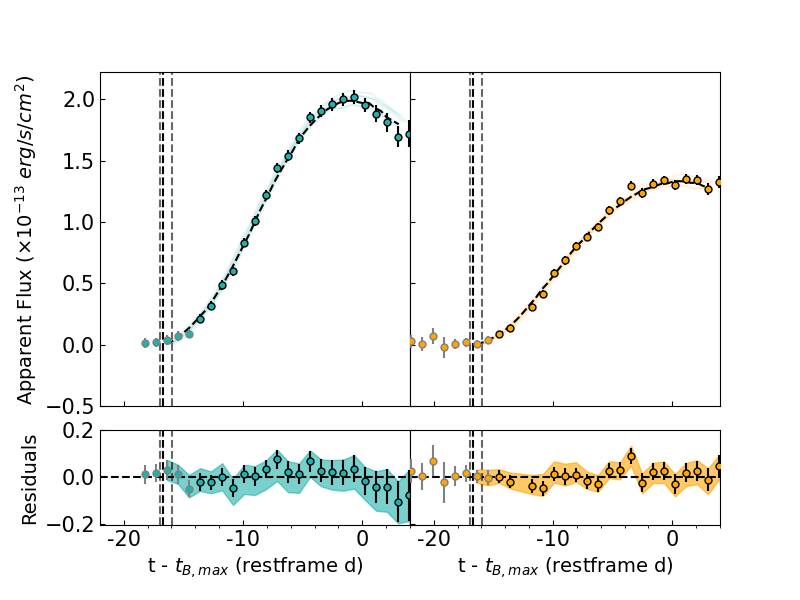}
\end{subfigure}%
\caption{In the top panels we show the flux \textit{g}-band (left) and \textit{r}-band (right) light curves of ZTF18abtnlik, which was well fit by models from the Chandrasekhar-mass model grid. The light curve is plotted in the rest frame of the SNe, where $t_{B, max}$ is taken from SALT2 fits performed by \protect\cite{Yao2019}. The models that fit within a 3-$\sigma$ range are shown as the coloured shaded lines, and the best-fitting model is represented by the black dashed line. Any $\leq 3 \sigma$ data points are included in the fitting routine, but have a grey outline to highlight the amount of data that would be lost if we fitted in magnitude space where these points would be considered upper limits. The range of estimated explosion dates within 3-$\sigma$ are shown by the grey dash-dot lines, and the explosion date of the best-fitting model is indicated by the vertical black dash-dot line. The residuals are also shown in flux in the same units as the top panel. The shaded region indicates the 3-$\sigma$ region of the best fit models and also accounts for the uncertainty on the explosion epoch as well as the photometric uncertainty. }\label{good_fits}
\end{figure}

We find that 78 out of 115 observed SN~Ia light curves (67~per cent) can be well reproduced by the \textsc{turtls} Chandrasekhar-mass model grid, implying that more than half of the variation observed in normal SN~Ia light curves can be reproduced by simply varying the parameters (\nick\ mass, \nick\ distribution, shape of the density profile, kinetic energy) in Chandrasekhar-mass explosions. Tables \ref{sn_table1} and \ref{sn_table2}  summarise the parameters of all SNe Ia in our sample, and and the light curves are shown in Figs. \ref{well_fit1}, \ref{well_fit2}, and \ref{well_fit3}. As discussed in Section \ref{comp_description}, the conditions for being `well-fit' are $\chi^2$ <= 4.7, an uncertainty on the explosion epoch of <= 3.3 d, and <= 22 unique matches in the 3-$\sigma$ range. In Fig.~\ref{good_fits} we present a sample light curve that is very well reproduced by our models.

\begin{figure}
    \centering
    \includegraphics[width=8cm]{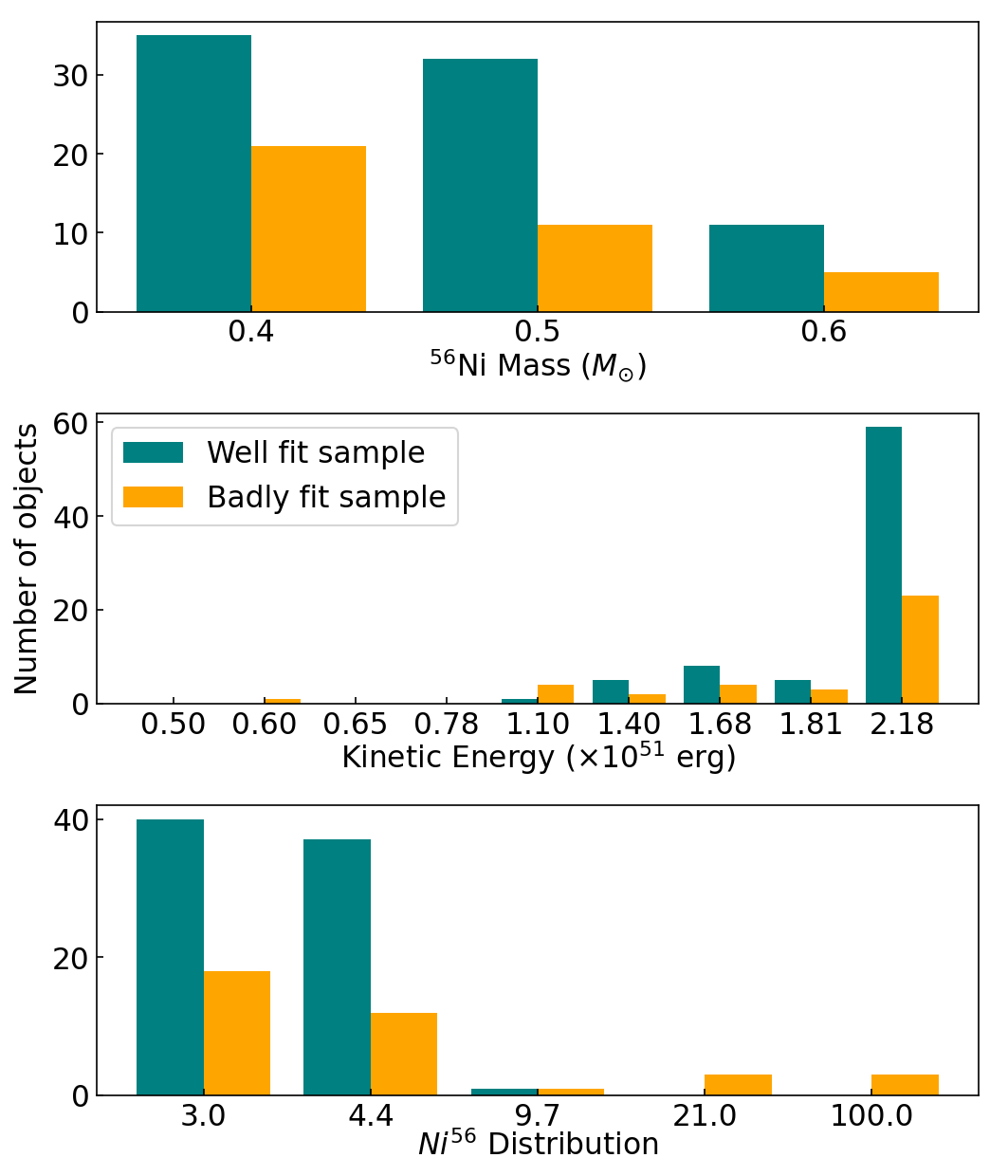}
    \caption{Histograms showing the distributions of $^{56}$Ni mass (top), kinetic energy (middle), and  \nick\ distribution in the ejecta (low values indicate a very extended distribution) (bottom) of the top match to each object in our well fit sample consisting of 78 objects (teal), and the badly fit sample consisting of 37 objects (orange). It is clear that the ZTF SNe~Ia sample shows a preference for highly energetic models with very extended \nick\ distributions. We also find that the objects matched with compact \nick\ distributions are all in the badly fit sample.}
    \label{distribution_parameters}
\end{figure}

Fig.~\ref{distribution_parameters} shows the distribution of the \nick\ mass, \nick\ distribution in the ejecta and kinetic energy of the well-fit and badly-fit objects. The density distribution of our well-fit sample is dominated by exponential profiles; only 18 well fit objects (23~per cent) have a double power-law density profile. The best-fitting $^{56}$Ni masses are distributed between 0.4--0.6 $M_{\odot}$ with a preference for lower masses. No objects were matched with an ejected $^{56}$Ni mass of 0.8 $M_{\odot}$, although it should be noted that these values are lower limits. Moreover, the typical magnitude difference at peak between our models with different $^{56}$Ni mass is $\sim$0.5 mag and given that the distance modulus can vary by up to 0.3 mag, there is some overlap between the best fit values for our SN~Ia sample. All well fit objects have more than one $^{56}$Ni mass in the 3-$\sigma$ range, and we therefore estimate the uncertainty on our $^{56}$Ni mass estimates to be $\pm$ 0.1 $M_{\odot}$. Models with highly extended $^{56}$Ni distributions dominate the sample: 77 out of 78 well fit objects are best matched by a model with $P$ = 3, 4.4 (the two most extended \nick\ distributions), with a single object being matched by $P$=9.7. Fig. \ref{distribution_parameters} shows that all SNe~Ia matched by compact \nick\ distributions are classified as badly-fit. Finally, models with a high kinetic energy dominate the sample (82~per cent, have ejecta kinetic energies >$1.4 \times 10^{51}$ erg). The preference for models with higher kinetic energies is currently not understood, and we are unable to determine whether this is an artifact of the fitting routine or an intrinsic property of the sample. There is some degeneracy between the $^{56}$Ni mass and the kinetic energy, which could be the cause of this skewed distribution, meaning these two parameters should be interpreted with caution. However, the \nick\ distribution is the main driver of the shape of the rise and is not heavily affected by the other parameters. 

\begin{table}
\begin{center}
\caption{The Pearson's and Spearman's correlation coefficients ($r$ and $\rho$ respectively) and their respective $p$-values for the model outputs (\nick\ mass and \nick\ distribution, $\exp$(\textit{P})) compared to the light curve observables (absolute peak \textit{g}-band magnitude, $M_g$, and $x_1$ parameter from SALT2 fitting). We check correlations with $\exp$(\textit{P}) rather than \textit{P} due to its formal definition in eq. \ref{equation}. We have highlighted in bold any significant correlations (those with correlation coefficients $\geq$ 0.50 and $p$-values < 0.05). }
\begin{tabular}{c|c|c|l|c|l}
\hline
\hline
     Parameter 1 & Parameter 2 & $r$ & $p$-value & $\rho$ & $p$-value\\
     \hline
    \nick\ mass  & $\exp$(\textit{P})  & 0.05 & 0.678 & -0.09 & 0.415\\
    \nick\ mass  & $x_1$      &  0.40 & $3\times 10^{-4}$ &  0.42 & $1\times 10^{-4}$\\
    \nick\ mass  &         $M_g$ & \textbf{0.70} & $2\times 10^{-12}$&  \textbf{0.70} & $4\times 10^{-12}$\\
           $\exp$(\textit{P})  &              $x_1$ &  0.06 & 0.577 & -0.07 & 0.519\\
           $\exp$(\textit{P})   &           $M_g$ & 0.08 &0.498 &  0.18 & 0.107\\
              $x_1$  &           $M_g$ & \textbf{0.51} & $2\times 10^{-6}$ & \textbf{0.52} & $9\times 10^{-7}$\\
\hline
\hline
\label{tab:corr_coefficients}
\end{tabular}
\end{center}
\end{table}

We investigated the correlations between the observed light curves properties \citep[absolute peak \textit{g}-band magnitude, $M_g$, and $x_1$ parameter from the SALT2 fitting performed by][]{Yao2019} and the model \nick\ masses and distributions for the 78 well-fit events. Table \ref{tab:corr_coefficients} shows a summary of the correlation coefficients. Unsurprisingly, $M_g$ is positively correlated with the \nick\ mass and $x_1$ value is correlated with $M_g$, in agreement with the well-known correlation between peak absolute magnitude and light curve shape \citep{Pskovskii1977, Phillips1993, Hamuy1995, Riess1996}. $x_1$ is also weakly correlated with \nick\ mass. Our sample is heavily dominated by objects with very extended \nick\ distributions in the ejecta (very low \textit{P} values), but due to the coarseness of the model grid in this parameter it is difficult to extract any trends. However, we encourage a future analysis with an increased grid resolution for the \nick\ distributions in order to further explore potential correlations with the shape of the rise.

\subsection{Badly-fit SNe~Ia}\label{badfittext}

\begin{figure*}
    \centering
    \includegraphics[width=15cm]{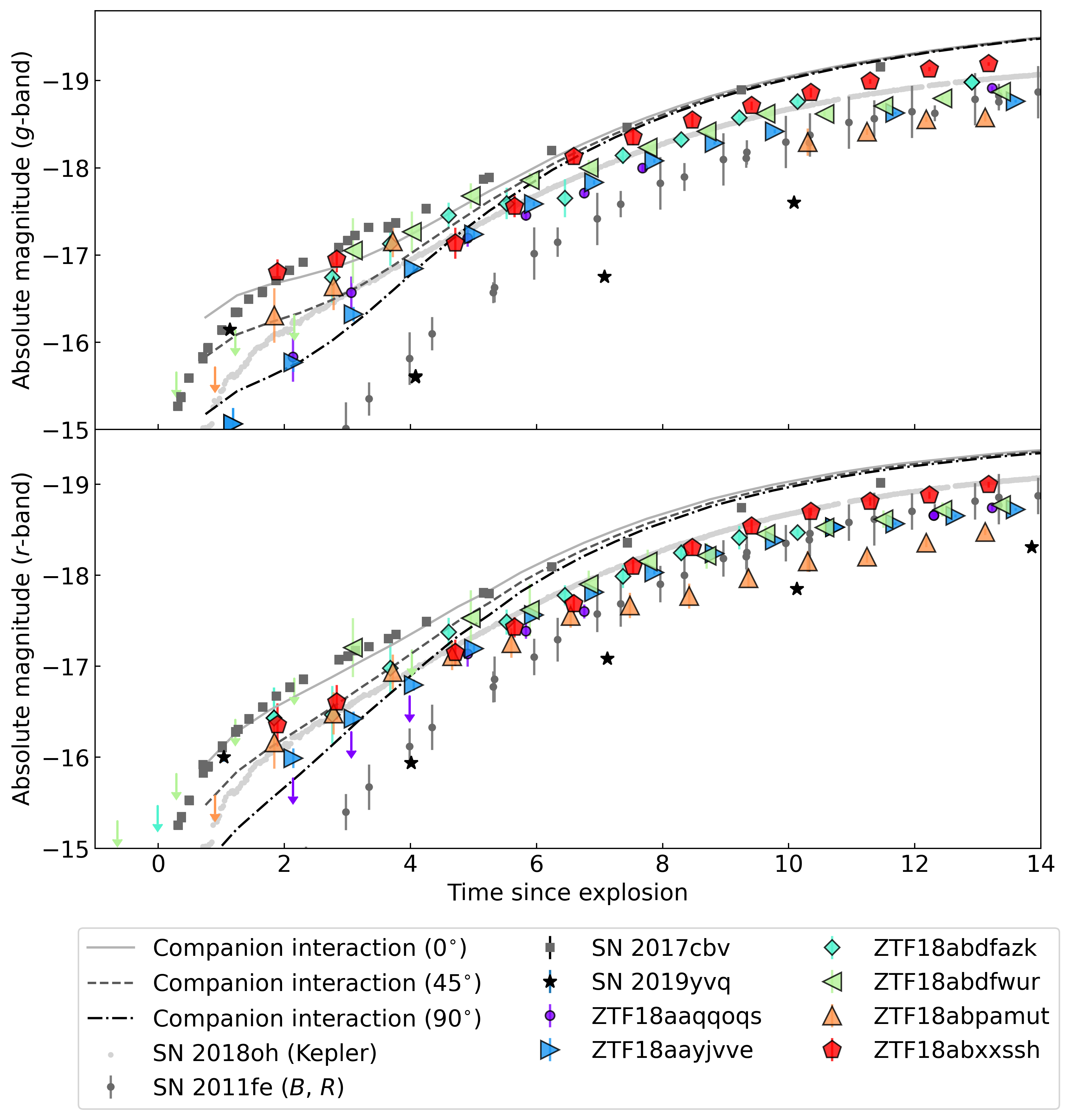}
    \caption{Plots showing the \textit{g}- (top panel) and \textit{r}-band (bottom panel) light curves of the six SNe Ia in our sample in magnitude space with a detected flux excess compared to three known SNe Ia with early flux excesses (SN 2017cbv, SN 2018oh, and SN 2019yvq) and the normal SN 2011fe which shows no flux excess. The Kepler light curve of SN 2018oh is plotted in both panels because \textit{gr}-band data was unavailable. For SN 2011fe, $B$-band and $R$-band data are shown in the upper and lower panels respectively. Upper limits in magnitude space are shown as downward pointing arrows, although we note that all fitting is done in flux space as detailed in Section \ref{bump_detection}. The companion interaction models used for the efficiency analysis are also shown, where all three models have a separation of $2.00\times 10^{12}$ cm but the models have various viewing angles ($0\degree$, $45\degree$, $90\degree$) resulting in flux excesses of differing strengths. Photometric data, distance moduli, and explosion epochs for SN 2017cbv, SN 2018oh, SN 2019yvq, SN 2011fe were taken from \protect\cite{Hosseinzadeh2017}, \protect\cite{Dimitriadis2019}, \protect\cite{Shappee2019}, \protect\cite{Miller2020b}, \protect\cite{Nugent2011}, and \protect\cite{Bloom2012b}.}
    \label{bumps_text}
\end{figure*}

For 37 out of 115 SNe~Ia (32~per cent), the model fits did not pass the cuts defined in Section \ref{comp_description}. Three objects (ZTF18abdfwur, ZTF18abpamut, and ZTF18abxxssh - see Fig. \ref{bumps_text}) fail to pass the cuts because they show a significant flux excesses at early times which is inconsistent with any of the models. Further analysis of these light curves is presented in Section~\ref{excess}. ZTF18abdfazk, ZTF18aayjvve, and ZTF18aaqqoqs also have a detected flux excess, but they pass the cuts and are included in the well-fit sample.

The other 34 excluded objects (See Fig. \ref{bad_fit1} for their light curves) fail the cuts for different reasons; eight fail because they have $\chi^2 \geq 4.7$, and seven fail only because the uncertainty on their explosion epoch is $\geq$3.3 d, with nineteen failing on more than one criterion. It is important to note that for objects with noisy data, which were ruled out only due to their number of unique matches or large uncertainty on the explosion epoch (21 SNe~Ia, 18~per cent) we are unable constrain their true explosion parameters, or conclusively determine whether they could be matched by Chandrasekhar mass explosions. We classify 13 light curves as "borderline" by eye (ZTF18aaqcugm, ZTF18aatzygk, ZTF18aawjywv, ZTF18aaxdrjn, ZTF18abfhryc, ZTF18abjtgdo, ZTF18abjtger, ZTF18abkhcrj, ZTF18abkhcwl, ZTF18abkifng, ZTF18abkudjo, ZTF18abqjvyl, and ZTF18abrzrnb). These objects have fits that generally look acceptable but marginally fail to pass the unique match cut of $\leq$22, or uncertainties on the explosion epochs are >3.3~d. As discussed in Section \ref{comp_description}, the cuts on the light curves that were applied are based on the distributions of the values for the overall sample and are somewhat arbitrary, but allow us to select the well-fit sample in a consistent and reproducible way. 

Only 13 out of 115 (11~per cent) which have a $\chi^2 \geq 4.7$, could not be matched by our models. Interestingly, seven of these have SALT2 $c$-value > 0.15. Out of the 115 objects in our sample, only ten have $c$ > 0.15. One of the other three with $\chi^2 < 4.7$, ZTF18abcysdx, is ruled out because the explosion date range spans 5.4 d and it has 29 unique matches, implying the parameters could not be adequately constrained. Of the two well fitted objects with $c$ > 0.15, ZTF18aaydmkh passes our cuts and falls into the well fit sample, but extinction and light curve shape appear to be an issue at late times, and it has $\chi^2$ = 2.5, which passes the cuts but is significantly higher than the mean of the sample. ZTF18abssdpi similarly passes the cuts, but with a range on the explosion dates of 2.9 d it lies near the edge of the distribution. To test whether extinction could be the cause of the shape mismatch for these objects, we corrected all SNe~Ia with $c$ > 0.15 for extinction \citep[values taken from][]{Bulla2020} and re-ran the fitting routine. This did not improve the model fits, and we conclude that the \textsc{turtls} models are not suited to modelling very red SNe~Ia, and any future analysis implementing \textsc{turtls} models should consider limiting their sample to objects with a SALT2 $c$-value < 0.15.

\subsection{SNe~Ia with early flux excesses}\label{excess}

\begin{figure}
    \centering
    \includegraphics[width=8.5cm]{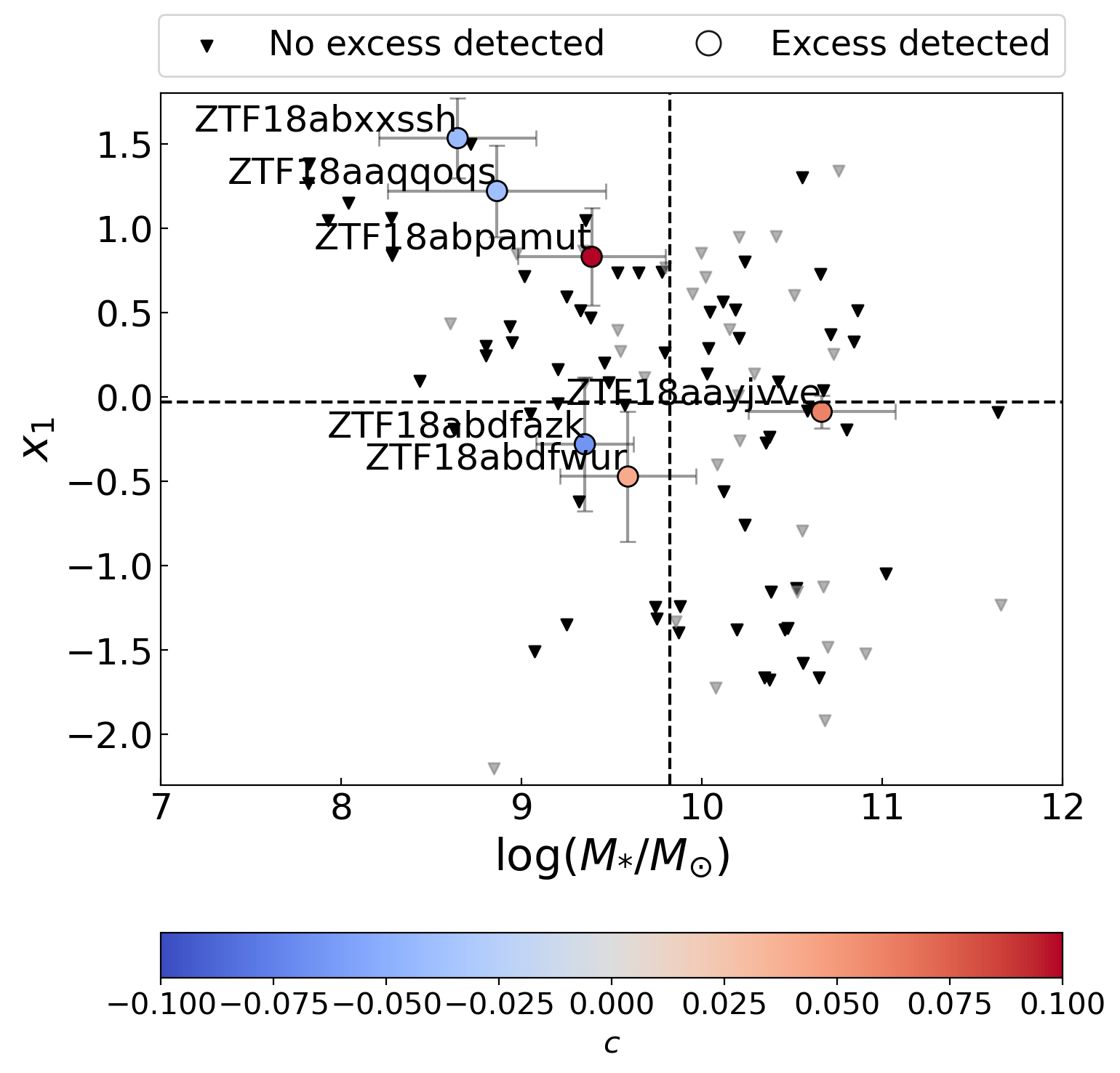}
    \caption{Plot of light curve stretch, $x_1$, as a function of the stellar mass of the host galaxy associated with each object. The colour indicates the $c$-value from the SALT2 fit. The black/grey triangles represent well fit/badly fit objects for which no flux excess was detected. The vertical dashed line shows the mean host galaxy mass of our sample (log($M_*$\slash$M_{\odot}$)=9.82), used here to show a division between low and high mass galaxies. The horizontal dashed line indicates the mean $x_1$ value of our sample (-0.03).}\label{x1_galaxy_mass}
\end{figure}

Using the method described in Section~\ref{bump_detection}, we identify six objects as having an early (within 6 days of explosion) flux excess relative to our underlying explosion models. Their \textit{g}- and \textit{r}-band light curves are shown in Fig.~\ref{bumps_text}. Three companion interaction models of \cite{Kasen2010} that are implemented in a flux excess efficiency analysis in Section \ref{efficiency} are also shown, along with the light curves of three literature events with prominent early flux excesses, SNe 2017cbv, 2018oh, and 2019yvq \citep{Hosseinzadeh2017, Shappee2019, Miller2020b}, and the light curve of the normal SN~2011fe \citep{Nugent2011, Bloom2012b}. Four of our objects (ZTF18abdfazk, ZTF18abdfwur, ZTF18abpamut, ZTF18abxxssh) have flux excesses with similar strengths to those of SN 2017cbv and 2018oh, while ZTF18aaqqoqs and ZTF18aayjvve have excesses that are less prominent.  All the flux excesses detected in our sample occur later than that of SN 2018oh (although we note that the comparison light curve for SN 2018oh is in the Kepler band which is roughly equivalent to a broad \textit{g+r+i} filter). None shows a flux excess that clearly declines and then rises again as was seen in SN 2019yvq but ZTF18abxxssh does show a plateau in its \textit{g}-band light curve at the earliest epochs.

The properties (spectral type, peak magnitude, light curve fit parameters, excess lifetime, significance of flux excess, \textit{g-r} colour at 3 d post explosion) of the six SNe Ia showing early flux excesses are given in Table \ref{bump_details}. We use the \textit{g-r} colour of the flux excesses at 3 d post the estimated explosion epoch as a way to determine if the early flux excess is intrinsically more prominent in the `blue' or `red'. We find that four of the early flux excesses have blue colours and two have red colours at 3 d post explosion. 
In Fig.~\ref{x1_galaxy_mass} we show the SALT2 light curve properties of $x_1$ and $c$ as a function of the host galaxy mass for all SNe~Ia in our sample with the objects displaying an early flux highlighted. Five out of the six SNe Ia with an identified early flux excess occur in a host galaxy with a mass lower than the mean of the sample and three of the six events have $x_1$ values in the top 20 percent of the sample. However, we are limited by the small number of SNe~Ia in the flux excess sample, and it is not possible to draw any statistically significant conclusions. There is an even split in the maximum-light colours ($c$) with three events having red colours and three having blue colours.

All the objects with a detected flux excess have a 3-$\sigma$ observation in the \textit{g}- or \textit{r}-band within 3~d of explosion. We show in Section \ref{efficiency} that the efficiency of detecting flux excesses drops off rapidly with increasing redshift, and the epoch of first detection scales with redshift. Early detections are therefore required to resolve the flux excess, but we also note that the presence of a flux excess will make it more likely that a SN~Ia is detected earlier. \cite{Yao2019} identified two objects in the ZTF 2018 sample with an early excess, ZTF18aavrwhu and ZTF18abxxssh. ZTF18abxxssh was identified by our code (e.g.~Fig. \ref{bumps_text}), whilst ZTF18aavrwhu is not picked up by our method because it can be well matched by Chandrasekhar-mass explosion models in our grid. \cite{Bulla2020} highlighted a further four objects in their analysis of a subset of the ZTF 2018 sample, which have a rapid colour transition (`red bumps') at early times: ZTF18abcflnz, ZTF18abcrxoj, ZTF18abpaywm, and ZTF18abgxvra. These SN~Ia do not show any form of flux excess when compared to our model grid but since they were identified based on their colour evolution it is not surprising that their sample does not overlap with ours.

\begin{table*}
\begin{threeparttable}
    \centering
    \caption{Summary of the properties of SNe~Ia with a detected flux excess.} 
    \label{bumps_table}
    \begin{tabular}{|l|c|l|l|c|c|c|c|c|c|c|c|c|c}
    \hline
        Name &Spectral type\tnote{a} & Peak mag. & $z$ & $x_1$&$c$& Lifetime of  & Ratio of excess & \textit{g-r} colour &\\
        &&&&& excess (d) & to peak flux\tnote{b} & at 3 d\tnote{c} \\
        \hline
ZTF18abdfazk& normal& $-19.2$& 0.084 & $-0.28$&$-0.065$&	5.5$\pm$0.7 &	0.073$^{\textrm{+}0.043}_{-0.021}$ & $-$0.28$\pm$0.43\\
ZTF18abdfwur& normal &$-18.9$& 0.070 & $-0.47$&0.041&		3.7$\pm$0.7&	0.111$^{\textrm{+}0.050}_{-0.061}$& 0.15$\pm$0.49\\
ZTF18abpamut &normal&$-18.6$& 0.064 &  0.83 &0.103&		6.6$\pm$2.4&	0.069$^{\textrm{+}0.037}_{-0.026}$& $-$0.16$\pm$ 0.36\\
ZTF18abxxssh &91T-like$^{d}$	&$-19.4$ &  0.064 &1.53& $-0.017$	&2.3$\pm$1.1&	0.067$^{\textrm{+}0.055}_{-0.016}$&	$-$0.35 $\pm$0.24\\

ZTF18aaqqoqs* &99aa-like  &$-19.3$& 0.082 & 1.22	& $-0.014$ &4.6$\pm$0.7 & 0.042$^{\textrm{+}0.027}_{-0.0001}$& 	$-$0.30$\pm$0.44\\
ZTF18aayjvve*  &normal& $-18.8$&  0.0474 & $-1.48$& 0.060 &2.4$\pm$0.5& 0.022$^{\textrm{+}0.012}_{-0.003}$& 0.10$\pm$0.11	\\
\hline
    \end{tabular}
    \label{bump_details}
    \begin{tablenotes}
    \footnotesize 
         \item[a]Spectral classification from template fitting. 
        \item[b]The ratio of the peak flux excess to the flux at maximum light. This ratio must exceed 0.02 in order for the SN~Ia to be categorised as having a flux excess.
        \item[c]Measurement of the \textit{g-r} colour at 3 d post the estimated explosion epoch. This is not necessarily during the peak of the flux excess, but all SNe~Ia with an identified flux excess have data at this epoch.
        \item[d]We have updated this classification from `normal SN Ia' from \protect \cite{Yao2019} based on the closer spectroscopic similarity to over-luminous 91T-like events.
        \item[*]ZTF18aaqqoqs and ZTF18aayjvve have flux excesses that are less significant than those previously detected in the literature. Therefore, to be conservative we do not include them in our intrinsic rate calculation. See Section \ref{efficiency} for further discussion.
    \end{tablenotes}
    \end{threeparttable}
\end{table*}

\subsubsection{Flux excess detection efficiency} \label{efficiency}

In order to determine the efficiency of our flux excess detection method, we use simulated light curves with early excesses based on the companion interaction models of \cite{Kasen2010} and attempt to recover them using the same detection method as for the data. We chose three models which reasonably resemble the range of flux excesses detected in our sample (Fig.~\ref{bumps_text}) but note that there are differences in the lifetimes and strengths of the observed bumps as a function of time that are not completely captured. Our choice of models does not imply a preferred physical origin for the flux excesses but is purely to test the detection efficiency of early flux excess detection. The models were produced by Magee et al. (subm.) using the companion interaction formulations presented in \cite{Kasen2010} combined with \textsc{turtls} SN Ia light curves (M20). The underlying Chandrasekhar-mass model is chosen to have values resembling our well-fitting sample (\nick\ mass of 0.5 \msun, kinetic energy of $1.68\times 10^{51}$ erg), and a \nick\ distribution with a $P$-value of 9.7 because more extended distributions result in a blended flux excess that would not be detected. All three companion interaction models have a radius of $2 \times 10^{12}$~m between the WD and the companion, and we choose viewing angles of 0$\degree$, 45$\degree$, and 90$\degree$, resulting in a range flux excesses of decreasing strength, which roughly covers the distribution of flux excesses found in the ZTF sample.

Next, we ensure the model light curves resemble the ZTF data. We separate the observed ZTF light curves with z $\leq$ 0.1 into three redshift bins of equal volume ($z$ $\leq$ 0.07, 0.07 < $z$ $\leq$ 0.087, 0.087 < $z$ $\leq$ 0.1). The bins sizes were chosen to ensure each bin contains sufficient SNe Ia to sample from (the highest redshift bin has 9 objects, which is the lowest number of SNe Ia across the bins). We draw photometric uncertainties from these redshift bins and apply these to the model light curves. The simulated light curves are initially setup at a one day cadence but to account for weather and other losses, we adjust each simulated light curve to have a cadence matched to a randomly drawn light curve from the observed sample. For ease of computation, the same number of data points are dropped in the \textit{g}- and \textit{r}-bands for an individual simulated event. We produce 15,000 simulated light curves, with equal contributions from each companion interaction model, spread evenly across the three redshift bins. We next apply our detection method to these simulated light curves in order to calculate the detection efficiencies for each redshift bin.

\begin{figure}
    \centering
    \includegraphics[width=8.5cm]{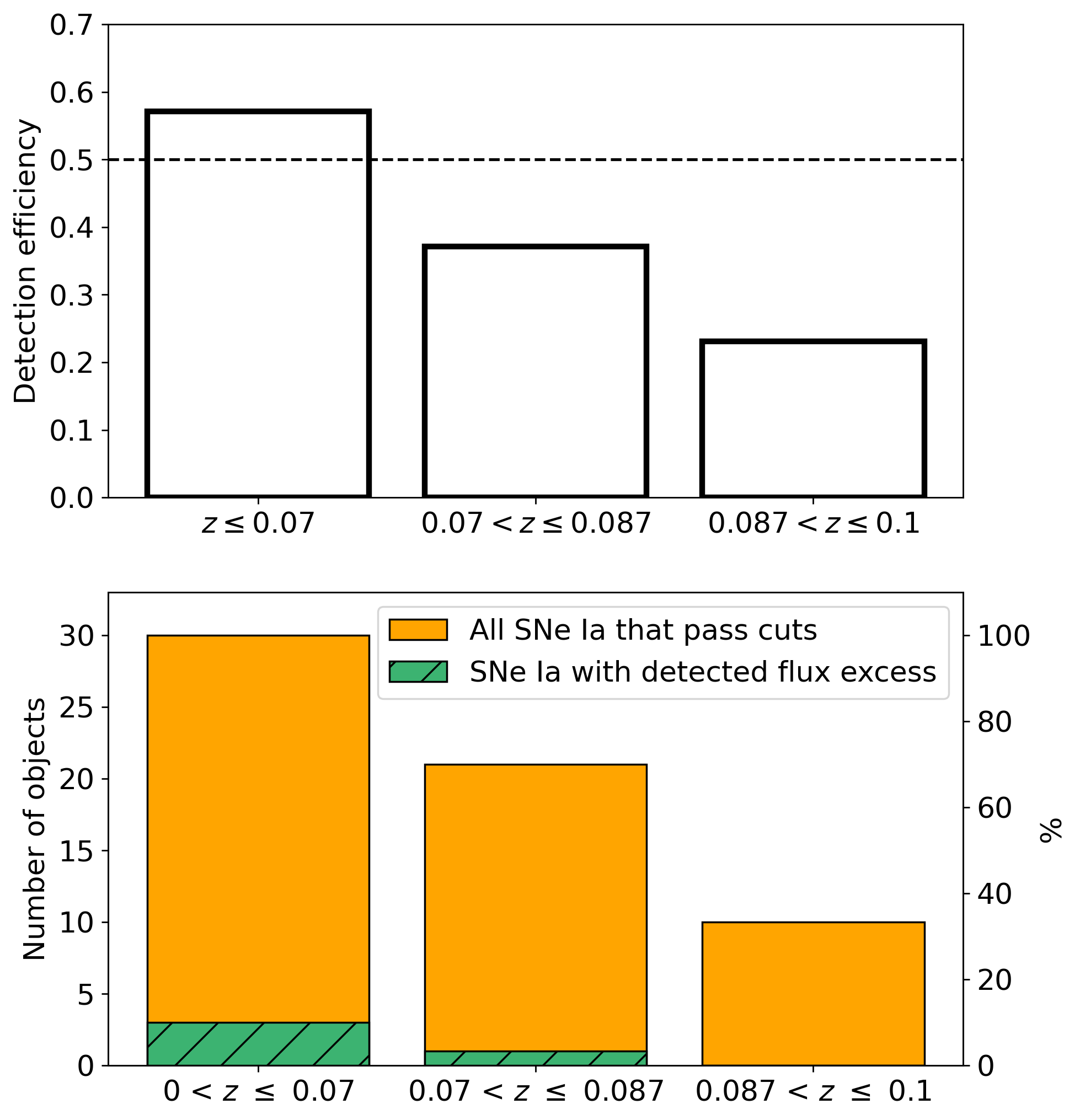}
    \caption{\textit{Top:} Histogram showing the detection efficiency of the flux excesses for three redshift bins for detections in \textit{g}- and/or \textit{r}-band. The efficiencies are based on 10,000 simulated light curves produced by combining the formulations of \protect\cite{Kasen2010} for companion interaction with viewing angles 0$\degree$ and 45$\degree$, with underlying Chandrasekhar mass model light curves (Magee et al., subm.) using typical uncertainties from the ZTF sample in each redshift bin. The efficiency drops below 50 per cent in the second redshift bin (0.07 < $z \leq$ 0.087, black dashed line). These efficiencies represent only the strongest flux excesses in the sample, and are based on the two strongest flux excess models. \textit{Bottom:} Histogram showing the number of observed ZTF SNe~Ia in our sample per redshift bin, and the number of those that have a detected flux excess (as a percentage of each bin on the right axis).}
    \label{efficiency_plot}
\end{figure}

We obtain efficiencies of 68 and 46 per cent in the lowest redshift bin for the strongest (0$\degree$ viewing angle) and middle (45$\degree$ viewing angle) model light curves, respectively. These detection efficiencies decline with increasing redshift as expected. For the weakest model light curve (90$\degree$ viewing angle), we find very low detection efficiencies ($\sim1$ per cent).  This model was included to account for the two weakest flux excesses in our sample (ZTF18aaqqoqs and ZTF18aayjvve). However, the efficiency is too low to perform rate calculations for these objects. There are two possible explanations for why we observed these weak flux excesses even though the efficiencies deem this unlikely, a) these two flux excesses are not real, or b) the weakest flux excess model is not representative of our weakest flux excesses. Fig. \ref{bumps_text} shows that these SNe~Ia have a clear excess in comparison to SN 2011fe, and therefore, we find the latter option more likely. This model light curve with the weakest early flux excess differs from the two weakest flux excess objects in both the excess duration as well as the peak luminosity from which the size of the flux excess is scaled. A further full parameter study, including a more detailed survey simulation \cite[e.g.][]{Feindt2019}, is needed to accurately determine the efficiencies of these lower significance flux excesses but is beyond the scope of this work. Therefore, we exclude these two events and the model with the weakest flux from further discussion of the efficiencies.

Fig.~\ref{efficiency_plot} shows the combined \textit{g}- and/or \textit{r}-band efficiencies as a function of redshift for the two brightest models. We find the highest detection efficiency (57$\pm$10 per cent) in the lowest redshift bin, with decreasing efficiency towards the highest redshift bin. We find that flux excesses are most frequently identified in the \textit{g}-band (50 per cent), and in half of these cases we also detect the excess in the \textit{r}-band. The flux excess is detected solely in the \textit{r}-band in 10 per cent of cases.

\section{Discussion}\label{discussion}

In this section, we first discuss the results presented in Section \ref{wellfittext} in the context of previous investigations of $^{56}$Ni masses and $^{56}$Ni distributions in Section~\ref{comparison}. In Section~\ref{implications}, we discuss the implications of our results on potential explosion mechanisms of SNe~Ia, by comparing the light curves of popular explosion models to those produced by our best-fitting Chandrasekhar-mass models. In Section \ref{excess_rates} we discuss our estimate of the intrinsic rate of flux excesses and compare to literature estimates of the rates of early flux excesses. In Section \ref{excess_discussion}, we present a summary of the properties of the sample of events with early excesses and discuss these in the context of previous observational and theoretical studies.

\subsection{Constraints on \protect\nick\ masses and distributions from early light curves}\label{comparison}

\begin{figure}
    \centering
    \begin{subfigure}{8cm}
    \includegraphics[width=8cm]{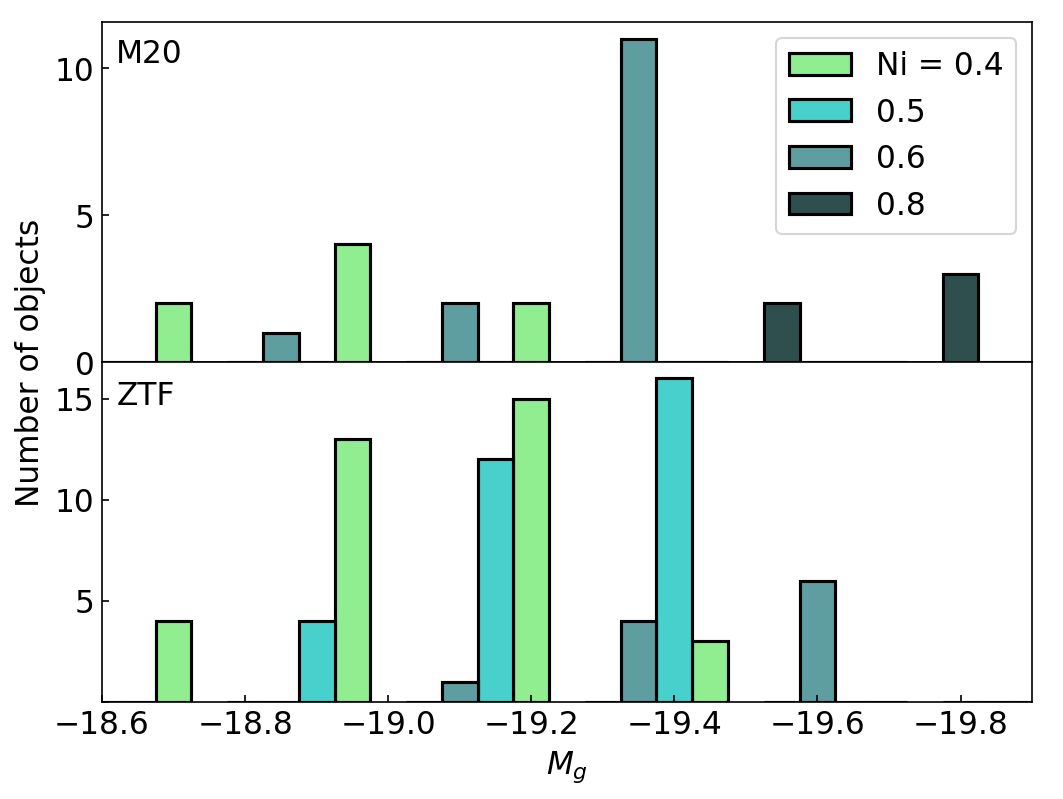}
    \end{subfigure}\vspace{0.5em}
    \begin{subfigure}{8cm}
    \includegraphics[width=8cm]{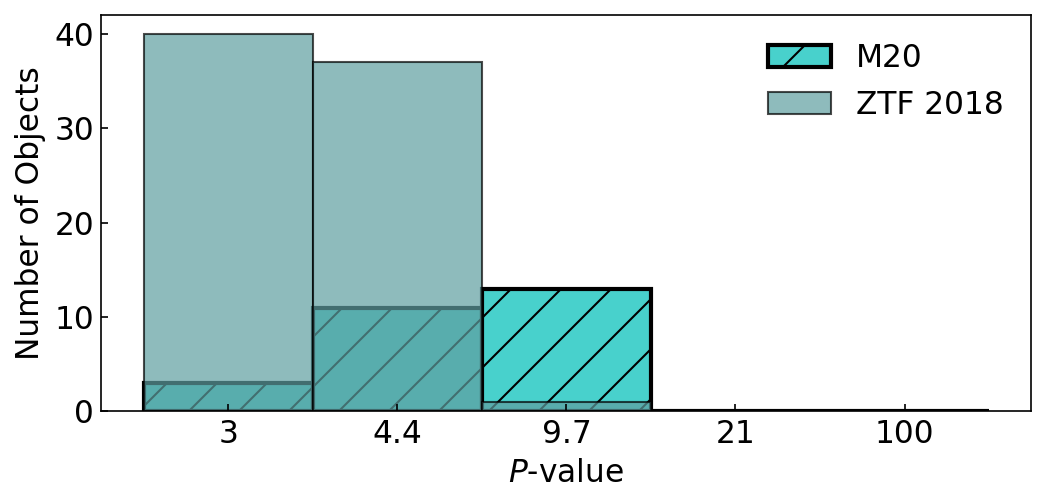}
    \end{subfigure}
    \caption{\textit{Top:} The stacked distribution of absolute magnitudes (\textit{g}-band) of the sample in M20, colour coded by the fitted ejected \protect\nick\ mass. The panel below shows the distribution of peak magnitudes of the ZTF 2018 sample colour coded by fitted \protect\nick\ mass. This figure shows that there are four objects which pass our cuts that have a peak magnitude (\protect$\sim$-19.6 mag) that could correspond to a \protect\nick\ mass of 0.8 \protect$M_{\odot}$ when compared to the results from M20, but are matched by 0.6 \protect$M_{\odot}$ in our analysis. We note that there are models in the grid that reach -19.6 mag with 0.6 \protect$M_{\odot}$, and moreover the model grid used for fitting in M20 did not contain models with \protect\nick\ = 0.5~\protect$M_{\odot}$, rendering a direct comparison more complicated. \textit{Bottom:} Histogram showing the distribution of \textit{P}-values of the well-matched objects in M20 compared to our sample. It is clear that the ZTF sample is skewed towards lower \protect\textit{P}-values, and therefore more extended \protect\nick\ distributions than the sample presented in M20. Nevertheless, both works agree that no normal SN~Ia light curve can be reproduced by a model with a highly compact \protect\nick\ distribution (\protect\textit{P} = 21, 100).}\label{comparing_peaks}
\end{figure}

We find that 67~per cent of SNe~Ia in our sample are well-matched by a Chandrasekhar-mass model. This is similar to the 74~per cent found by M20 for a literature sample of 35 SNe~Ia, in particular when considering that for 18 per cent of objects we were not able to constrain the parameters but are unable to definitively rule out a Chandrasekhar mass model. We further find that ZTF SNe~Ia are best matched by models with $^{56}$Ni masses distributed between 0.4--0.6 $M_{\odot}$ (Fig.~\ref{distribution_parameters}), whereas the objects in M20 are distributed across the 0.4--0.8 $M_{\odot}$ range, peaking at 0.6 $M_{\odot}$. To understand this difference, we compare the peak magnitude distribution of both samples in Fig.~\ref{comparing_peaks}. We find that the literature sample in M20 has three objects in the brightest magnitude bin, where the ZTF sample has zero. This is likely driven by selection effects: it is easier to identify flux excesses in brighter SNe~Ia at earlier phases than fainter ones, and thus brighter objects get published more. Since the peak magnitude is correlated with the $^{56}$Ni mass, it is not surprising that the M20 matches several SNe~Ia with a $^{56}$Ni mass of 0.8 $M_{\odot}$. Moreover, the model grid implemented in M20 does not include models with \nick\ = 0.5 $M_{\odot}$, impeding a direct comparison between the distributions. 
Our $^{56}$Ni mass range of 0.4--0.6 $M_{\odot}$ is in general agreement with the large sample (337 SNe\,Ia) analysed in \cite{Scalzo2014a}, where they found that the majority of SNe~Ia have $^{56}$Ni masses of 0.3--0.7 $M_{\odot}$. 

Our sample of SNe~Ia is dominated by model light curve with highly extended $^{56}$Ni distributions ($P$ = 3, 4.4), whereas the SNe~Ia analysed by M20 also match models with more compact $^{56}$Ni distributions ($P$ = 9.7, Fig.~\ref{comparing_peaks}). We can not identify a clear reason for this difference in $^{56}$Ni distributions between the literature sample and the ZTF sample. The ZTF SN~Ia sample should be relatively unbiased in its selection, while the M20 sample is based on individual objects or small samples of objects that have been published independently. There is a clear difference in the peak absolute magnitudes, and associated $^{56}$Ni masses (Fig.~\ref{comparing_peaks}), for the two samples. Since the ZTF sample is larger and lower bias, we take this to be a better representation of the SNe~Ia population than the literature sample of M20, and conclude that the $^{56}$Ni distributions of SNe~Ia that are well matched by Chandrasekhar-mass explosion models appear to be highly extended.

We perform a comparison between the rise times computed from the \textsc{turtls} models (measured from explosion to maximum light) and those computed by \cite{Miller2020} (measured from first light to maximum light), and find these to be generally consistent. In approximately half of cases the rise time computed from the \textsc{turtls} models are more than 1~d longer than those computed by \cite{Miller2020}, consistent with the existence of a dark phase. However, due to the breakdown of the LTE assumption around maximum light, the estimates of peak light from the \textsc{turtls} models have large uncertainties. The uncertainty on the explosion date and coarseness of the model grid further contribute to the uncertainty in the estimation of the rise time. Similarly, the rise time presented in \cite{Miller2020} rely on a conversion from the peak in the \textit{g}-band to \textit{B}-band, introducing non-negligible uncertainties. For these reasons, we do not present any further analysis on the duration of the dark phases.

\cite{Bulla2020} analysed the colour evolution of 65 SNe~Ia from the ZTF 2018 data set that had a colour measurement within 5~d of first light. They found that the colour evolution of most objects was relatively flat and covered the full range of $^{56}$Ni distributions presented by M20 implying that at least some degree of extended $^{56}$Ni is required, but some objects can also be matched by the most compact distributions ($P$ = 21, 100), which disagrees with both our findings and the results presented in M20. As noted by M20, a more realistic composition structure would shift colours to redder values meaning that caution must be taken when comparing the absolute colours of the models to observations.

\subsection{Implications of extended \nick\ distributions for SN Ia explosion mechanisms} \label{implications}

\begin{figure*}
    \centering
    \includegraphics[width=14cm]{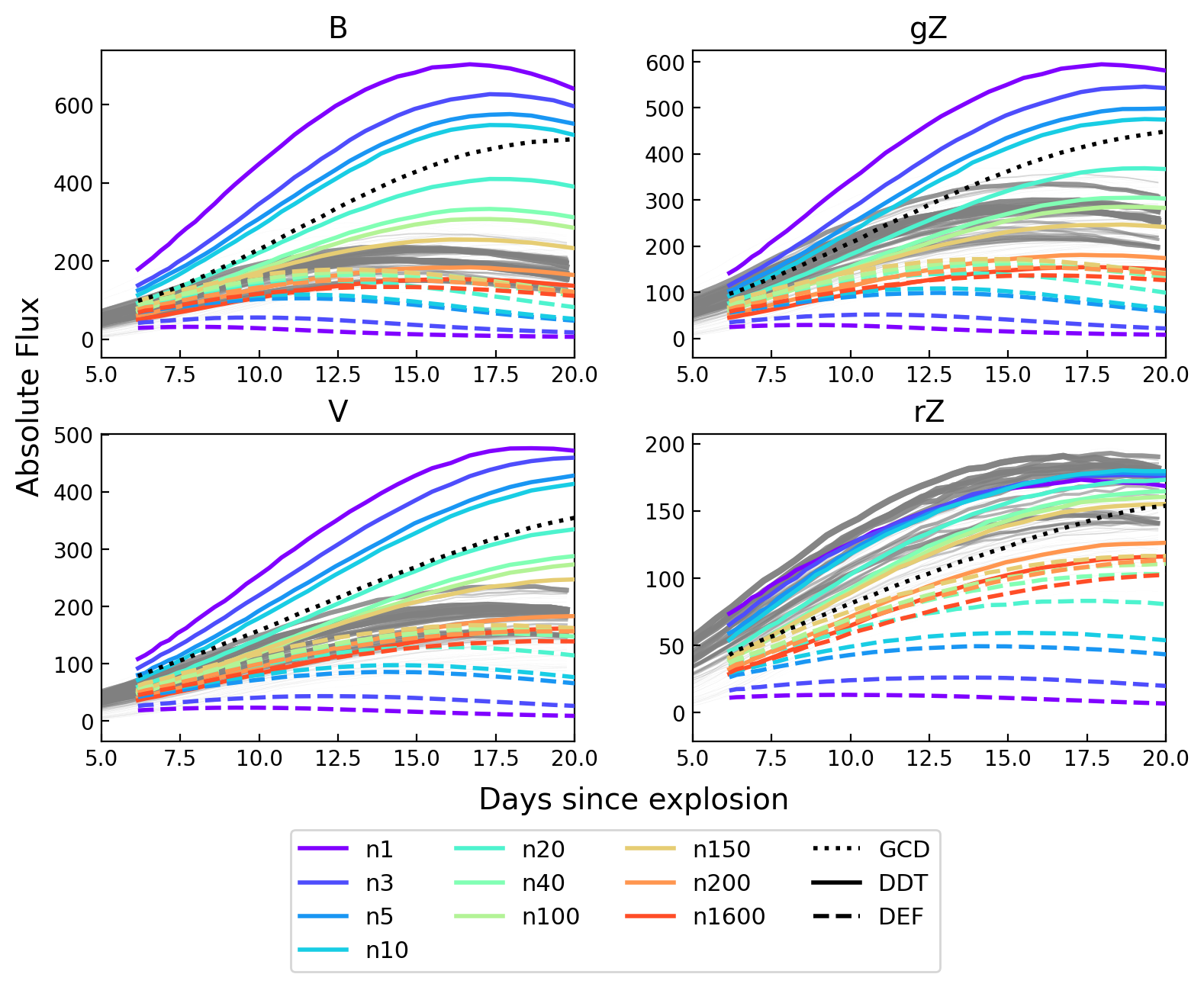}
    \caption{Plots showing the comparison of the \textsc{turtls} light curves to DDT \citep{Seitenzahl2013}, DEF \citep{Fink2014}, and GCD models \citep{Seitenzahl2016}. Only \textsc{turtls} models that were matched to objects in our well fit sample are plotted (grey), where the line width indicating the number of times that particular model was matched by a SN~Ia. The plots show the light curves in \textit{B}, \textit{V} (Bessel) and the \textit{g}, \textit{r} (ZTF) bands, for ingition kernels ranging from $n$=1 to $n$=1600. While DDT models produce light curves that are well matched by our models, the DEF models are unable to reproduce the shape of the rise and the peak magnitude.}\label{model_lc_comparison}
\end{figure*}

In this section, we discuss the preference for highly extended $^{56}$Ni distributions in the ZTF SNe~Ia sample in the context of SNe~Ia explosion models. In Fig.~\ref{model_lc_comparison}, we compare the light curves of models that were matched to SNe~Ia in our sample to theoretical light curves for DDT \citep{Seitenzahl2013}, deflagration \citep[`DEF'][]{Fink2014}, and GCD models \citep{Seitenzahl2016}. Fig.~\ref{model_lc_comparison} shows that the DDT models produce light curves that are well matched by the \textsc{turtls} models, specifically those with 40-1600 ignition kernels (\texttt{n40 - n1600}), where the \texttt{n100} is generally taken to be representative of normal SNe~Ia \citep[][]{Seitenzahl2013}. The higher fraction of $^{56}$Ni in the outer regions found in our objects also resembles the DDT models, although these models show a steeper decline in the $^{56}$Ni mass fraction near the outer layers (see fig.~1 in M20), which results in a more rapid rise in the early light curve. By delaying the time between the detonation and deflagration, more mixing could bring the DDT models in better agreement with \textsc{turtls} models with the most extended \nick\ distributions. As noted by M20, however, this could have a significant effect on the ejecta composition, and so further modelling is required to determine which configurations of the DDT models would best reproduce light curves from our model grid with the lowest \textit{P}-values.

It has been shown that DEF models struggle to reproduce the light curve characteristics (peak magnitude and colour) and distinctive spectral features (Si, S and Ca) seen in normal SNe~Ia \citep{Fink2014}. We also find that the DEF models are unable to reproduce the shape of the rise and the peak magnitudes reached by the \textsc{turtls} models. DEF models are a better match to 02cx-like SNe (SNe~Iax), a subclass easily distinguished from normal SNe~Ia by their faint magnitudes and low ejecta velocities \citep{Kromer2013, Fink2014, Leung2020a}. Since the deflagration models often leave behind a bound remnant, the ejected $^{56}$Ni masses are significantly lower \citep[0.035 - 0.315 $M_{\odot}$;][]{Fink2014}, complicating a direct comparison of the absolute magnitudes reached between these models. Nonetheless, the shape of the rising light curve has not been directly compared to a large sample of SNe~Ia, and we show that deflagration models result in a flatter rise than the range set by the SNe~Ia from the ZTF sample.

The GCD model describes the explosion of a near-Chandrasekhar mass WD, and the light curve evolution of this model is also shown in Fig.~\ref{model_lc_comparison}. Although this model produces a $^{56}$Ni distribution more comparable to \textsc{turtls} models with $P$ = 3, GCD models tend to produce large amounts of $^{56}$Ni \citep[0.74 $M_{\odot}$;][]{Seitenzahl2016} which is reflected in the high peak magnitudes of the GCD model in Fig.~\ref{model_lc_comparison}. In addition to being over luminous, the GCD models are expected to produce highly asymmetrical ejecta and since SNe~Ia generally show little polarisation indicative of a symmetric explosion geometry \citep{Bulla2015}, the GCD models are unlikely to be the dominating explosion mechanism for normal SNe~Ia. 

\cite{Scalzo2014a} analysed the ejected mass distribution of SNe~Ia and suggested that sub-Chandrasekhar mass WDs make a significant contribution to the SNe~Ia population. To rule out sub-Chandrasekhar mass WDs for the SNe~Ia in this sample, additional modelling would be required. Some models can be speculatively ruled out based on the distinct characteristics of their resulting light curves. Explosions of WD with a thick He-shell that burn to Fe-group elements produce a significant excess \citep{Jiang2017, Noebauer2017, Polin2019}, which would allow us to discern them from $M_{Ch}$ explosions. Furthermore, some He-shell detonation mass models predict an early red colour evolution as a result of the line blanketing from the IGE produced in the He-shell \citep{Nugent1997, Kromer2010, Woosley2011, Polin2019}, and since the majority of objects in the ZTF sample show a relatively constant colour evolution, apart from six which are identified by \citet{Bulla2020} as having a "red" bump, we speculate that a double-detonation is unlikely to be the dominating explosion mechanism for our well fit sample. However, some double-detonation models have also been shown to reproduce normal SNe~Ia properties, without producing a significant excess at early times \citep{Shen2018, Polin2019, Magee2021, Gronow2021b}. In particular, double-detonations with He-shells that do not burn to Fe-group elements can not be excluded \citep{Magee2021}.

\subsection{The intrinsic rate of flux excesses in SNe Ia} \label{excess_rates}

\begin{figure*}
\centering
    \includegraphics[width=13cm]{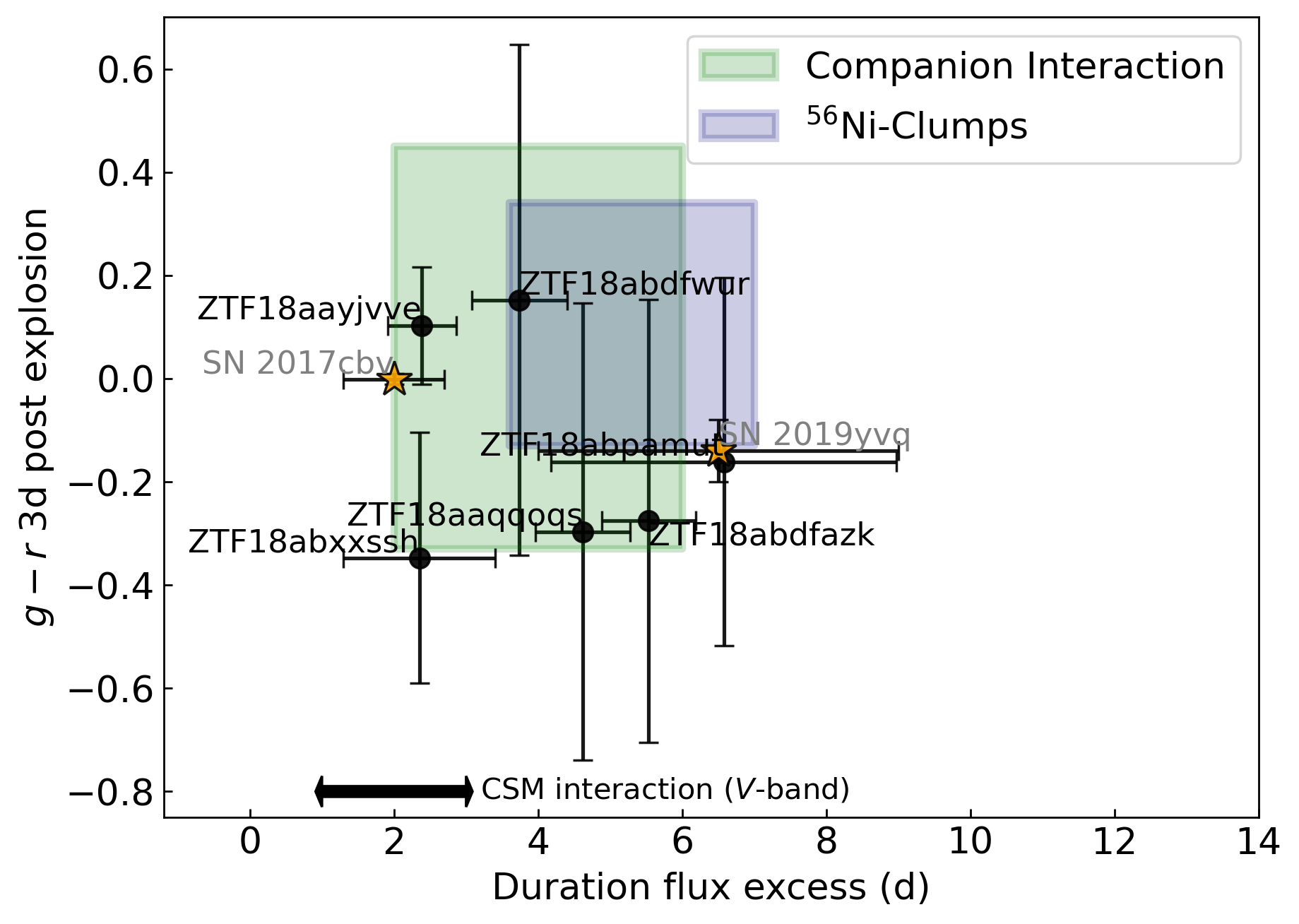}
    \caption{A plot showing the $g-r$ colour at 3 d post explosion of the SNe Ia with a detected flux excess, as a function of the duration of the flux excess. The shaded region show the parameter space occupied by companion interaction \protect\citep{Kasen2010, Magee2021} and \protect\nick\ -clump models \protect\citep{Magee2020a}. The range of lifetimes expected from CSM-interaction models produced by \protect\cite{Piro2016} are shown, but no $g-r$ colour predictions are made. The $g-r$ colour and flux excess lifetimes for SN 2017cbv and SN 2019yvq are also plotted. The durations of the flux excesses for four of the events (ZTF18aaqqoqs, ZTF18abdfazk, ZTF18abdfwur, ZTF18abpamut) appear to be longer than those predicted by the CSM interaction models of \protect \cite{Piro2016}.}\label{lifetimes}
\end{figure*}

We combine the detection efficiencies presented in Section \ref{efficiency}
with the observed rate of early flux excesses to estimate the intrinsic rate of events with flux excesses. As noted in Section \ref{bump_detection}, the observed rate is calculated by applying the same cuts on light curve scatter to the full sample as to the flux excess sample. We use only the lowest redshift bin ($z$ $\leq$ 0.07), which contains three SNe~Ia (out of 30 events) with a flux excess, to estimate the intrinsic rate of flux excesses. As discussed previously ZTF18aaqqoqs and ZTF18aayjvve are not included in the calculation of the intrinsic rate because we are unable to constrain the efficiency for these weaker flux excesses. We combine the \textit{g}- and \textit{r}-band flux excesses because all excesses are visible across both bands. The uncertainty on the intrinsic rate is derived by combining the uncertainty of the detection efficiency and the Poisson uncertainty on the number of detected flux excesses. 

For `normal' and 91T/99aa-like SNe Ia, we estimate an intrinsic rate of flux excesses of similar strength to those presented in the literature of 18 $\pm$ 11 per cent. This is consistent with the calculated intrinsic rate in the second redshift bin (0.07 < $z$ $\leq$ 0.087) of 12$\pm$19 per cent, which contains one SN~Ia with a flux excess out of 21 objects (observed rate of 5 per cent) in this bin. However, due to the efficiency estimate in this bin dropping below 50 per cent and only a single SN~Ia with a detected flux excess, we quote the value from the lowest redshift bin as the intrinsic rate. We note that due to the simplified efficiency calculation and small number of flux excess SNe~Ia in the sample, there are likely significant statistical and systematic uncertainties associated with this estimate of the intrinsic rate. A full survey simulation using e.g.~\textsc{simsurvey} \citep{Feindt2019} would be required for a more accurate estimate of the efficiency of flux excess detections, particularly at the lower significance end.

Our calculated intrinsic rate of 18 $\pm$ 11 per cent of SNe Ia showing a flux excess is similar to that of M20, who found a rate of 22~per cent in a sample of 23 literature objects. Our rate is lower than \cite{Jiang2018} who found a rate of 35 per cent in a sample of 23 objects, although it should be noted that their analysis is focused on characterising flux excesses rather constraining the rate of flux excesses. Since we have performed our analysis on the much larger unbiased year-one ZTF SN Ia sample first presented in \cite{Yao2019}, it is unsurprising that our rate is lower than previous estimates because of the untargeted nature of our sample. Moreover, the rates estimated by the above mentioned studies represent absolute observed rates, whereas our measurement is an intrinsic rate. \cite{Olling2015} perform an in-depth study of the high-cadence Kepler light curves of 3 SNe~Ia, and find no evidence of any flux excesses. In total, there are 4 Kepler light curves of nearby SNe~Ia \citep[KSN~2012a, KSN~2011b, KSN~2011c, SN2018oh;][]{Olling2015,Dimitriadis2019, Shappee2019, Li2019} with only one showing a flux excess (SN~2018oh), suggesting a rate of 25 per cent. However, we note the small number statistics for this sample.

\subsection{Potential origins for early flux excesses in the ZTF SN~Ia sample} \label{excess_discussion}

In Section \ref{excess}, we presented the SNe~Ia in the ZTF 2018 sample for which we detected a flux excess and their general light curve and spectral properties. In Section \ref{comparison_bumps_data_models}, we present an overview of the properties of the early flux excesses and the relation to progenitor scenarios, while in Section \ref{91t_discuss}, we discuss their broader properties and the potential link in some cases to the 91T/99aa-like subclass and younger stellar populations.

\subsubsection{Quantifying the diversity of early flux excesses}
\label{comparison_bumps_data_models}

Figure \ref{lifetimes} shows the $g-r$ colours at 3 d post explosion and lifetimes of the excesses of the 6 SNe~Ia with identified flux excesses. The duration of the bump is calculated assuming a start time of the bump as the median time between the last non-detection and the first non-zero detection of the residuals from the best-fitting model with an uncertainty on the start time of half the time between these points. The end time was calculated in a similar fashion. We also show the $g-r$ colours at 3 d post explosion and lifetimes of the excesses for the literature events, SNe 2017cbv and 2019yvq. 

We compared the flux excess sample to models producing a flux excess \citep[companion interaction, CSM interaction, and \nick\  clumps;][]{Kasen2010, Piro2016, Maeda2018a, Magee2020a, Magee2021}. The parameter space covered by the companion interaction and \nick-clump models are overplotted in Fig.~\ref{lifetimes}, along with the typical lifetime range for the CSM interaction models. For the companion interaction models, the interaction signature at early times is expected to produce blue emission since most of the emission from the shock is radiated in the UV \citep{Kasen2010}. However, for a realistic comparison with observed events, it must be combined with an underlying ejecta model. Magee et al. (subm.) implemented the analytical formulae from \cite{Kasen2010} to add a companion interaction component to an underlying Chandrasekhar-mass model light curve. These models resulted in a $g-r$ range of $-$0.3 -- $-$0.45 mag at 3 d post explosion and lifetimes of 2 -- 7 d. 

\cite{Piro2016} provided no direct $g-r$ colours for the CSM-interaction models, but they predict a $B-V$ colour range of 0.1 -- 0.4 mag during the interaction-powered flux excess. Assuming a blackbody-like spectrum at 3 d post explosion (near the end of the short duration CSM interaction), we expect the $g-r$ colours to be 0.25 mag redder but again are complicated by the contribution of the underlying ejecta \citep[see discussion in][]{Piro2016}. The lifetimes of flux excesses caused by CSM interaction are generally short, ranging from 1 -- 3 d. The \nick-clump models of \cite{Magee2020a} also provide predictions for the early colour and duration of any flux excess. For this scenario, we exclude models which produce $g-r$ < $-$0.5 mag near maximum light for being unrealistic matches for our sample with the remaining models predicting $g-r$ = $-$0.13 -- 0.34 at $\sim$ 3d post explosion and lifetimes of the flux excess of 3.6 -- 7.0 d. We note that these models were produced to match SN 2017cbv and SN 2018oh, and therefore these models only represent a sub-section of the full parameter space. Double-detonations models that burn to IGE in the shell predict a `red bump' in the early light curve as the result of line blanketing \citep{Polin2019, Bulla2020, Magee2021}. However, we do not include these predictions in Fig. \ref{lifetimes} because these models are also predicted to produce very red maximum-light spectra that are not seen for these events. Double-detonation models that burn to only IMEs look normal at maximum light but they are not predicted to have flux excesses at early times. 

We have found that the early flux excesses have a variety of strengths, colours and lifetimes (see Fig.~\ref{bumps_text} and Fig.~\ref{lifetimes}). The majority of the objects with a detected flux excess in our sample (apart from ZTF18abxxssh) do not show the distinct rise followed by a drop/plateau as seen in e.g.~SN 2019yvq \citep{Miller2020b} but four events (ZTF18abdfazk, ZTF18abpamut, ZTF18aaqqoqs, ZTF18aavrwhu) show flux excesses that are of a comparable strength to the literature objects, SN 2017cbv and SN 2018oh. ZTF18abxxssh shows the most prominent shoulder (Fig.~\ref{bumps_text}) at early times of the sample. All of our events with early flux excesses rise earlier and more rapidly than predicted by the underlying Chandrasekhar mass models, producing a shoulder in the light curve and therefore, likely need an external source of energy to produce them. 

The peak magnitude of the excess can range between 2 -- 11 per cent of the flux at maximum light (the lower limit is set by the detection method), have $g-r$ colours during the flux excess of $-$0.4 to 0.2 mag, and lifetimes of the excesses of 2 -- 7 d. For two of the events in our sample (ZTF18aayjvve, ZTF18abxxssh), the lifetimes of the bumps are short at $\sim2$ d and are most consistent with the predictions of the companion (or CSM) interaction models \citep{Kasen2010,Piro2016} and are inconsistent with the \nick\ clump model lifetime predictions. They fall in the region covered by the colours predicted for the companion interaction model (no $g-r$ colours are available for the CSM interaction models). For the other four events with flux excesses the lifetimes are inconsistent with the CSM interaction predictions but overlap in the colour and lifetime space of the companion interaction and \nick\ clumps models. Conclusively determining the cause of a flux excess based on the light curve without spectra obtained during the excess is difficult, although we are able to rule out CSM interaction as a cause for excesses lasting more than 3 d, corresponding to four events in our sample  (ZTF18abdfazk, ZTF18abdfwur, ZTF18abpamut, ZTF18aaqqoqs). 

We find that three (ZTF18abdfazk,  ZTF18abxxssh, ZTF18aaqqoqs) of the four SNe Ia displaying a blue colour at 3 d post explosion also have a negative $c$ value at maximum light, following the previous results for SNe 2017cbv and 2018oh that had bluer than average colours \citep{Hosseinzadeh2017,Dimitriadis2019,Li2019,Shappee2019}. ZTF18abxxssh and ZTF18aaqqoqs also have very broad light curves and maximum-light spectra consistent with overluminous SNe Ia (see \ref{91t_discuss} for further comparison with such events). The two events (ZTF18abdfwur, ZTF18aayjvve) with red early colours also show red colours at maximum light but not extreme values. An interesting conclusion of our model and data comparisons of the colour of the early flux excesses is that although the physical processes involved in the production of the flux excesses may produce hot blue emission, this needs to be combined with the underlying SN ejecta component that can result in an overall redder colour. As shown in Fig.~\ref{lifetimes}, the parameter space of companion interaction and \nick\ clump model predictions with an Chandrasekhar-mass model ejecta covers relatively red, as well as blue, early light colours. This suggests that studies should be careful to not exclude early flux excess scenarios that predict blue colour solely because the observed colour is redder. Spectra are key for distinguishing between the origins of flux excesses and rapid spectroscopy should be a focus of future flux excess studies.

\subsubsection{A connection to young stellar populations?}
\label{91t_discuss}

The sample size is very small but as demonstrated in Section \ref{excess}, the SNe~Ia with flux excesses in our sample have a preference for lower stellar mass hosts (Fig.~\ref{x1_galaxy_mass}). This suggests that these objects originate from younger stellar populations \citep{Rigault2020}, which is likely more consistent with the SD scenario. Alternatively, since lower mass galaxies tend to host brighter SNe~Ia \citep[e.g.~those with broader light curves;][]{Sullivan2010}, it is easier to detect brighter SNe~Ia at early times. However, our method measures the relative flux excess between the data and model, and assuming the luminosity of the flux excess does not scale with the luminosity of the SNe explosion, it should be easier to identify a bump in fainter SNe~Ia, or one with its \nick\ distribution more constrained towards the centre (high \textit{P} value). This is not evident from our small sample, which is populated by both faint and bright SNe Ia, making it unclear whether brighter, more energetic, SNe~Ia explosions produce more prominent bumps.

\cite{Jiang2018} performed a study of the light curves of SNe~Ia from the literature with early flux excesses and categorised them based on the shapes and evolution of their early light curves. They found that all three 91T-like events in their sample have strong flux excess in their early light lasting over a week and the three 99aa-like events in their sample have either broad early light curves, or flux excesses lasting up to a week. Of our six events with early flux excesses, we identify one object as a 99aa-like SN Ia (ZTF18aaqqoqs), and one as a 91T-like SN Ia (ZTF18abxxssh). In keeping with the typical characteristics of these classes, these two events have some of the broadest light curves (high $x_1$ values) of the sample and are found in lower stellar mass hosts than average. Their early bumps are also the two bluest of the sample and they also show blue colours at maximum light. However, the lifetimes of their early excesses differ with ZTF18aaqqoqs being significantly longer (4.6$\pm$0.7 d) than the flux excess lifetime of ZTF18abxxssh  (2.3$\pm$1.1 d). Therefore, it is unclear if they have the same origin for their early flux excesses but their overlap in other properties is of note. 

We also wish to determine if we can confirm the result of \cite{Jiang2018} that all 91T/99aa-like SNe~Ia have some form of excess at early times. To do this we focus on the early light curves of SNe Ia in our full sample that are classified as 99aa/91T-like but were not identified by our method as having an early flux excess. We find ten such events. Seven of these (ZTF18aaumeys, ZTF18aaxakhh, ZTF18abixjey, ZTF18abklljv, ZTF18abmmkaz, ZTF18abpmmpo, ZTF18abrzrnb) have insufficient early data to detect a flux excess or the uncertainties in the early data are too large. One event (ZTF18abauprj) does not show a flux excess but is not well matched to any model at peak, which may affect the fit at early times. Two events (ZTF18aaytovs, ZTF18abfwuwn) have sufficient early time data to search for an early excess but show no excess suggesting that not all 99aa/91T-like events in an unbiased sample show an early excess. We can estimate a rough rate of 91T/99a-like events showing early flux excesses using the observed rates and our detection efficiencies. The lowest redshift bin (0 $< z <$ 0.07) of our sample contains four 99aa/91T-like SNe Ia (including our reclassification of ZTF18abxxssh). We identify one flux excess in this lowest redshift bin (ZTF18abxxssh), setting a lower limit on the observed rate of flux excesses for 91T/99aa like SNe Ia at 25 per cent. The intrinsic rate derived from the four 91T/99aa-like SNe Ia in the first redshift bin is 44$\pm$13 per cent when considering the detection efficiency (57 per cent) in this bin. This rate should be considered with caution since there is only one 91T/99aa-like SN~Ia with a detected flux excess in this redshift bin. We do not confirm the results found by \cite{Jiang2018} that 100~per cent of 91T/99aa-like SNe~Ia have some form of excess, but the frequency does appear to exceed that of normal SNe~Ia and we encourage future studies with larger samples of these sub-classes to investigate this in greater detail.

\section{Conclusions} \label{conclusion}

We fit 115 SNe~Ia light curves from the ZTF 2018 sample \citep{Yao2019, Miller2020} with a grid of Chandrasekhar-mass explosions \citep{Magee2018, Magee2020} with varying degrees of $^{56}$Ni mixing and find that approximately 67~per cent of objects can be fit by a model from our grid with \nick\ masses in the range of 0.4 to 0.6 \msun. We are unable to adequately constrain model parameters for 19~per cent of the sample, and only 11~per cent could not be matched by a Chandrasekhar mass model. We find that no light curves can be fit by the most compact \nick\ distributions ($P$= 21, 100), confirming the result found by M20, although our sample shows a stronger preference for the most highly extended distribution (lowest $P$-values). The best-fitting models from our grid are found to be generally consistent with the DDT of \cite{Seitenzahl2013}, although the steep decline towards the outer layers is not matched by the \textsc{turtls} models, and further explosion modelling is needed to explore this. 

We performed a search for the presence of flux excesses in the early light curves of the sample, and find 6 objects that display a flux excess.  We calculate detection efficiencies from a sample of 10,000 simulated light curves in the \textit{g}- and \textit{r}-bands. Based on three SNe~Ia with a flux excess out of 30 SNe~Ia (detected rate of 10 per cent) in the lowest redshift bin ($z \leq 0.07$), and a detection efficiency of 57 per cent in this redshift bin, we find an intrinsic rate of strong flux excesses in SNe Ia of 18$\pm$11 per cent. This is consistent with the intrinsic rate derived from the $0.07<z\leq~0.087$ redshift bin of 12$\pm$19 per cent, which contained one SN~Ia with a flux excess out of 21 SNe~Ia (5 per cent). We analysed the \textit{g-r} colours and lifetimes of the flux excesses, and find that these are generally are consistent with the ranges predicted by both interaction (companion or CSM) and \nick\ clump models, although longer lasting excesses are unlikely to be caused by CSM interaction. SNe Ia displaying a flux excesses tend to occur in lower mass galaxies. We could be missing flux excesses because the SNe~Ia are not detected early enough, which highlights the importance of finding infant SNe within 3~d of explosion. Moreover, flux excesses are predicted to peak in x-rays, and a strong signal is expected in the UV/NUV. The various models predict some of the signal to leak into the optical bands, enabling us to detect the flux excesses. However, with more early UV/NUV data, the detection efficiencies could be significantly improved, particularly since the contribution from the underlying SN Ia ejecta will be weaker in the UV/NUV allowing a cleaner detection.

We encourage future searches for early flux excesses with one day cadence surveys (e.g. ZTF-II), and UV/NUV surveys. This will allow us to detect SNe Ia flux excesses at a higher efficiency, and catching them earlier will enable us to trigger spectroscopic follow up. Rapid spectroscopic follow up is crucial and could provide us with the information needed to discern between the different origins of the flux excesses (e.g.~interaction with a companion or CSM, He-shell detonations, or \nick\ clumps in the outer ejecta). Further analysis of the properties of flux excesses provide a promising method to unveiling the progenitors and explosion scenarios of SNe~Ia.

\section{Acknowledgements}

The authors would like to thank the anonymous referee for helpful comments that have significantly improved this paper.

MD, KM, and MRM are funded by the EU H2020 ERC grant no. 758638. MR  (+ MS) has (have) received funding from the European Research Council (ERC) under the European Union’s Horizon 2020 research and innovation programme (grant agreement n°759194 - USNAC, PI: Rigault). MMK acknowledges generous support from the David and Lucille Packard Foundation. This work was supported by the GROWTH project funded by the National Science Foundation under Grant No 1545949. MWC acknowledges support from the National Science Foundation with grant number PHY-2010970. 

Based on observations obtained with the Samuel Oschin Telescope 48-inch and the 60-inch Telescope at the Palomar Observatory as part of the Zwicky Transient Facility project. ZTF is supported by the National Science Foundation under Grant No. AST-1440341 and a collaboration including Caltech, IPAC, the Weizmann Institute for Science, the Oskar Klein Center at Stockholm University, the University of Maryland, the University of Washington, Deutsches Elektronen-Synchrotron and Humboldt University, Los Alamos National Laboratories, the TANGO Consortium of Taiwan, the University of Wisconsin at Milwaukee, and Lawrence Berkeley National Laboratories. Operations are conducted by COO, IPAC, and UW. Funding for the Sloan Digital Sky Survey IV has been provided by the Alfred P. Sloan Foundation, the U.S. Department of Energy Office of Science, and the Participating Institutions. SDSS acknowledges support and resources from the Center for High-Performance Computing at the University of Utah. The SDSS web site is www.sdss.org. This work made use of the Heidelberg Supernova Model Archive (HESMA), \href{https://hesma.h-its.org}. This work used the zttps forced photometry tool for identifying potential targets \citep{Reusch2020}.
The Pan-STARRS1 Surveys (PS1) and the PS1 public science archive have been made possible through contributions by the Institute for Astronomy, the University of Hawaii, the Pan-STARRS Project Office, the Max-Planck Society and its participating institutes, the Max Planck Institute for Astronomy, Heidelberg and the Max Planck Institute for Extraterrestrial Physics, Garching, The Johns Hopkins University, Durham University, the University of Edinburgh, the Queen's University Belfast, the Harvard-Smithsonian Center for Astrophysics, the Las Cumbres Observatory Global Telescope Network Incorporated, the National Central University of Taiwan, the Space Telescope Science Institute, the National Aeronautics and Space Administration under Grant No. NNX08AR22G issued through the Planetary Science Division of the NASA Science Mission Directorate, the National Science Foundation Grant No. AST-1238877, the University of Maryland, Eotvos Lorand University (ELTE), the Los Alamos National Laboratory, and the Gordon and Betty Moore Foundation.

\section{Data Availability}
The \textsc{turtls} model grid implemented in this study is available at \url{https://github.com/MarkMageeAstro/turtls-Light-curves} and is described in detail in \cite{Magee2018}. All the ZTF light curves from the sample are described in \cite{Yao2019} and any public data is available on the ZTF GROWTH Marshall \citep{Kasliwal2019}. Full forced photometry light curves will be shared on reasonable request to the corresponding author. The simulated light curves produced for the efficiency analysis are available at \url{https://github.com/deckersm/ZTF2018_Deckers_Paper}.

\bibliographystyle{mnras}
\bibliography{mnras_template}

\appendix

\section{}
The \textit{g}- and \textit{r}-band light curves and the matched \textsc{turtls} models are shown in Figures \ref{well_fit1}, \ref{well_fit2}, \ref{well_fit3}, and \ref{bad_fit1}. We present the host galaxy data in Tables \ref{galaxy_data} and \ref{galaxy_data2}. A summary of the \textsc{turtls} fits, as well as the SALT2 parameters derived in \cite{Yao2019} are presented in Tables \ref{sn_table1} and \ref{sn_table2}.

\begin{figure*}
    \centering
    \includegraphics[width=16cm]{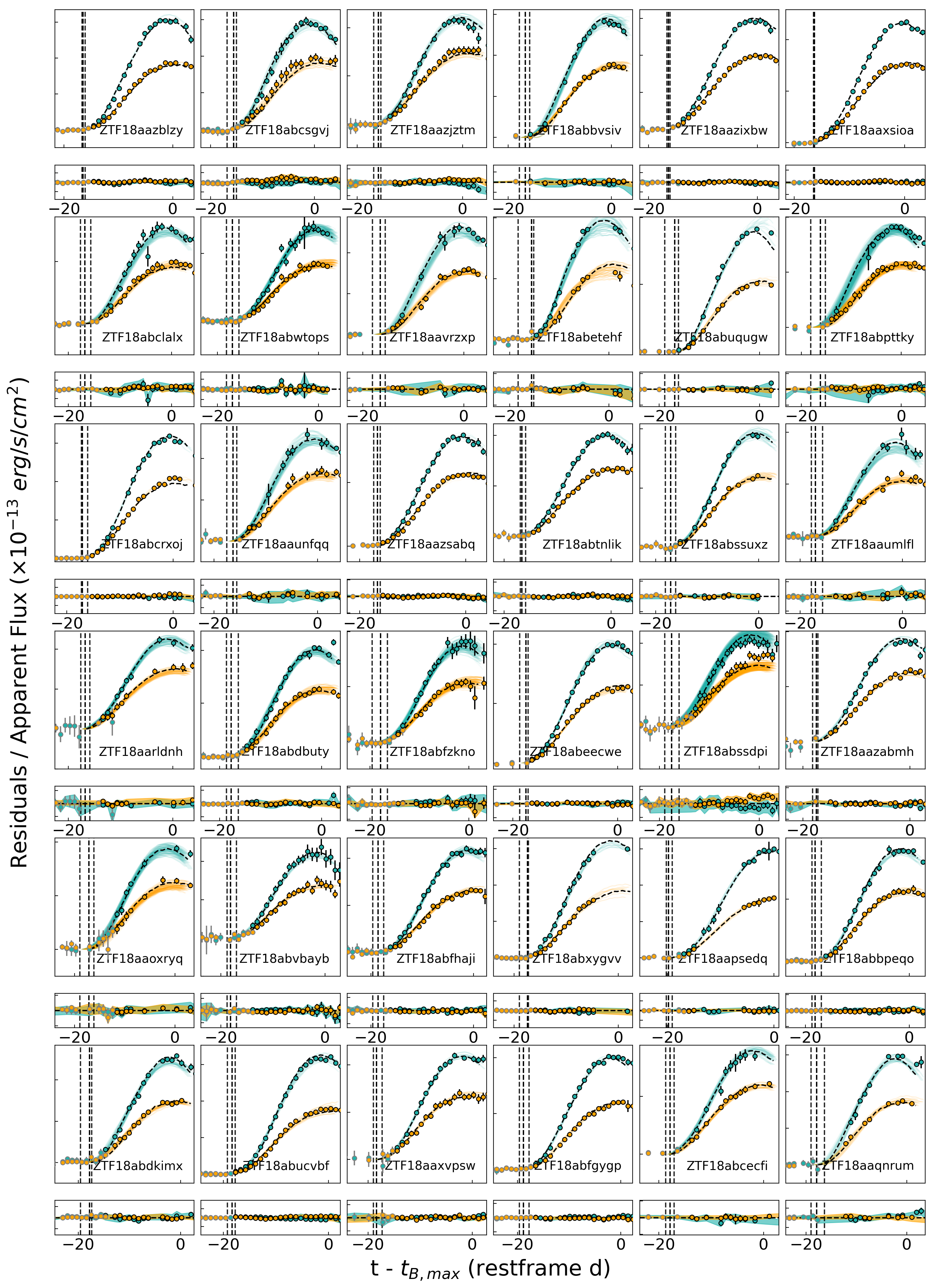}
    \caption{The light curves of the well fit SNe~Ia in the sample, ordered alphabetically.}
    \label{well_fit1}
\end{figure*}

\begin{figure*}
    \centering
    \includegraphics[width=16cm]{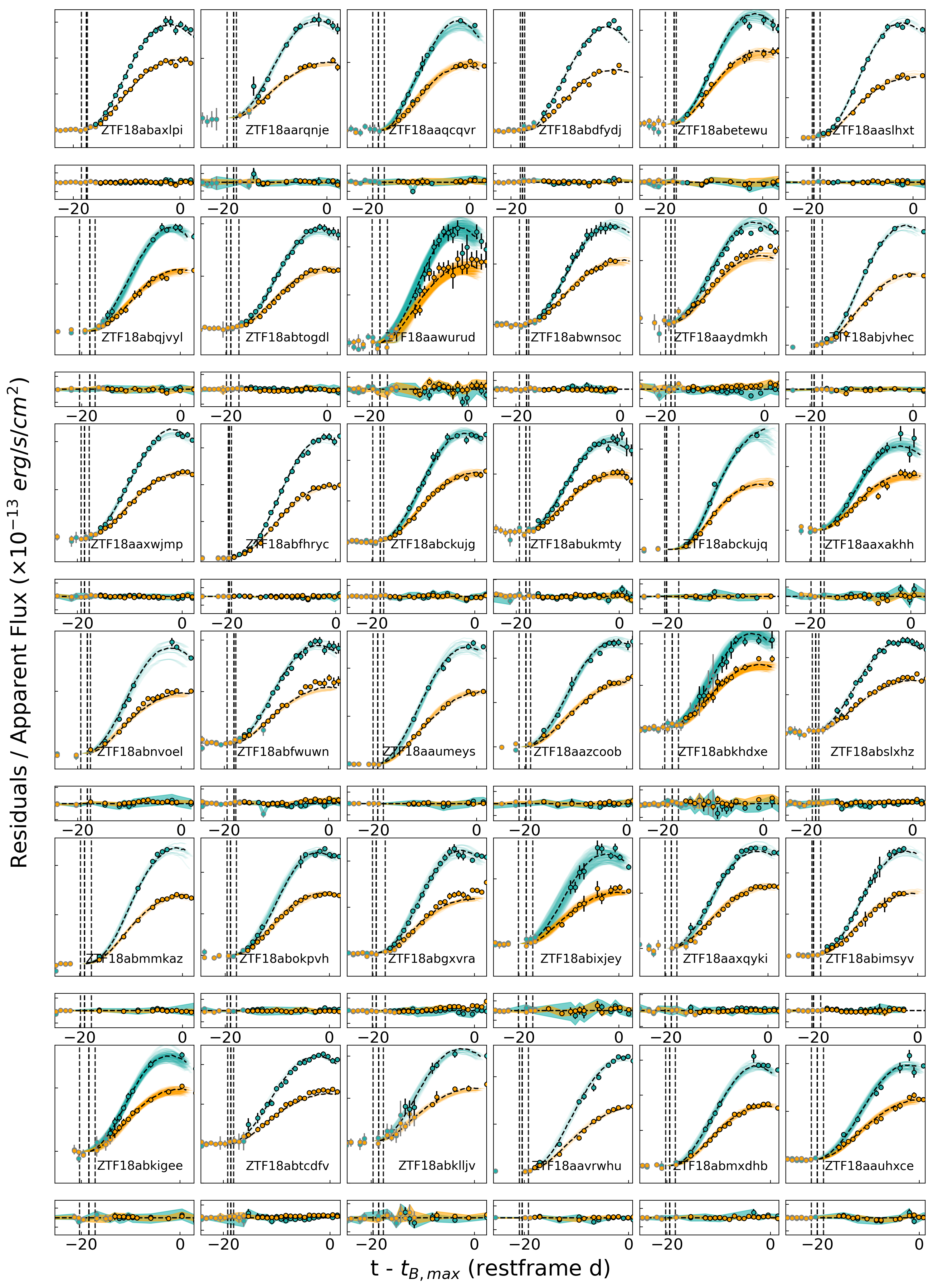}
    \caption{Continuation of Fig. \ref{well_fit1}}
    \label{well_fit2}
\end{figure*}

\begin{figure}
    \centering
    \includegraphics[width=8.5cm]{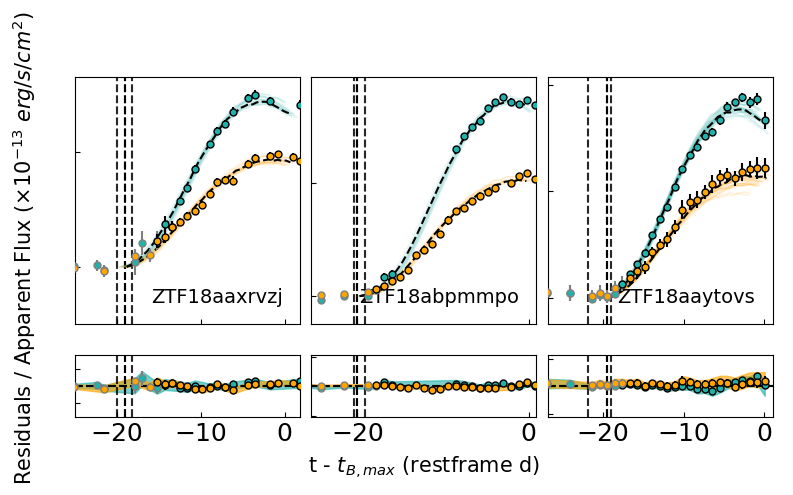}
    \caption{Continuation of Fig. \ref{well_fit1}}
    \label{well_fit3}
\end{figure}

\begin{figure*}
    \centering
    \includegraphics[width=16cm]{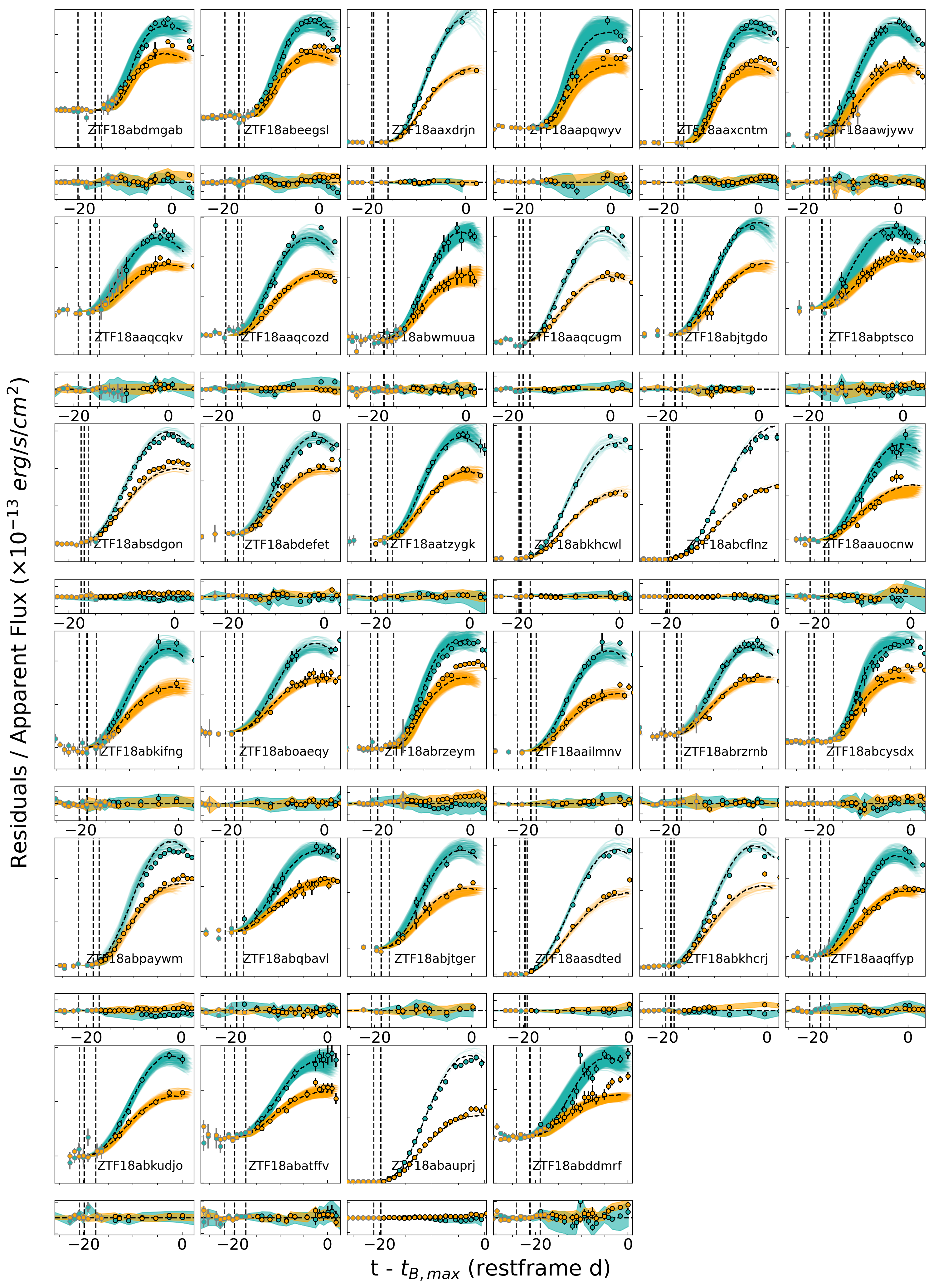}
    \caption{The light curves of the badly fit SNe~Ia in the sample, ordered alphabetically.}
    \label{bad_fit1}
\end{figure*}

\newpage
\begin{table}
    \caption{Values of the mean host galaxies stellar mass and uncertainties for the sample, as well as the DLR calculated for between each object and its identified host galaxy.}\label{galaxy_data}
    \centering
    \begin{tabular}{|l|l|l|}
    \hline
        Object &  Host Galaxy Stellar Mass & DLR\\
        \hline
ZTF18aailmnv&	10.150 $\pm$	0.319&	0.925\\
ZTF18aaoxryq&	10.374 $\pm$		0.351&	0.126\\
ZTF18aapqwyv&	10.077 $\pm$		0.339&	0.161\\
ZTF18aapsedq&	11.640 $\pm$		0.383&	1.913\\
ZTF18aaqcozd&	11.657 $\pm$		0.355&	2.851\\
ZTF18aaqcqkv&	9.854 $\pm$		0.503&	0.509\\
ZTF18aaqcqvr&	9.457 $\pm$		0.348&	1.190\\
ZTF18aaqcugm&	10.671 $\pm$		0.403&	5.718\\
ZTF18aaqffyp&	9.342 $\pm$		0.298&	1.810\\
ZTF18aaqnrum&	8.433 $\pm$		0.762&	4.221\\
ZTF18aaqqoqs&	8.863 $\pm$		0.605&	1.908\\
ZTF18aarldnh&	11.018	$\pm$	0.270&	1.590\\
ZTF18aarqnje&	9.202 $\pm$		0.546&	3.391\\
ZTF18aasdted&	8.969 $\pm$		0.385&	0.652\\
ZTF18aaslhxt&	10.034 $\pm$		0.186&	1.778\\
ZTF18aatzygk&	10.201 $\pm$		0.307&	0.182\\
ZTF18aauhxce&	10.555 $\pm$		0.227&	3.445\\
ZTF18aaumeys&	9.013 $\pm$		0.338&	1.877\\
ZTF18aaumlfl&	10.524 $\pm$		0.348&	2.626\\
ZTF18aaunfqq&	9.742 $\pm$		0.397&	0.773\\
ZTF18aauocnw&	10.288 $\pm$		0.403&	1.522\\
ZTF18aavrwhu&	-&		1.573\\
ZTF18aavrzxp&	10.460 $\pm$		0.422&	1.150\\
ZTF18aawjywv&	10.697 $\pm$		0.353&	0.524\\
ZTF18aawurud&	10.842	$\pm$	0.324&	0.395\\
ZTF18aaxakhh&	10.184	$\pm$	0.334&	0.774\\
ZTF18aaxcntm&	10.904 $\pm$		0.401&	0.800\\
ZTF18aaxdrjn&	10.679 $\pm$		0.326&	2.211\\
ZTF18aaxqyki&	- &		27.734\\
ZTF18aaxrvzj&	7.820 $\pm$		1.540&	1.550\\
ZTF18aaxsioa&	9.073 $\pm$		0.253&	1.277\\
ZTF18aaxvpsw&	10.671 $\pm$		0.391&	0.241\\
ZTF18aaxwjmp&	8.933 $\pm$		0.564&	0.262\\
ZTF18aaydmkh&	10.204 $\pm$		0.321&	0.378\\
ZTF18aayjvve&	10.663 $\pm$		0.408&	2.362\\
ZTF18aaytovs&	10.779 $\pm$		0.348&	0.936\\
ZTF18aazabmh&	- &		36.070\\
ZTF18aazblzy&	8.638 $\pm$		0.671&	16.070\\
ZTF18aazcoob&	10.654 $\pm$		0.361&	2.661\\
ZTF18aazixbw&	10.561 $\pm$		0.326&	1.421\\
ZTF18aazjztm&	10.345 $\pm$		0.347&	0.669\\
ZTF18aazsabq&	9.876 $\pm$		0.367&	0.786\\
ZTF18abatffv&	10.409 $\pm$		0.499&	2.577\\
ZTF18abauprj&	10.758 $\pm$		0.333&	3.989\\
ZTF18abaxlpi&	10.026 $\pm$		0.310&	0.887\\
ZTF18abbpeqo&	10.585 $\pm$		0.308&	1.204\\
ZTF18abbvsiv&	10.647 $\pm$		0.341&	1.303\\
ZTF18abcecfi&	10.421 $\pm$		0.366&	1.641\\
ZTF18abcflnz&	9.682 $\pm$		0.323&	3.049\\
ZTF18abckujg&	10.041 $\pm$		0.225&	2.529\\
ZTF18abckujq&	10.863 $\pm$		0.467&	2.731\\
ZTF18abclalx&	9.869 $\pm$		0.463&	1.559\\
ZTF18abcrxoj&	-&	1.376\\
ZTF18abcsgvj&	10.373 $\pm$		0.377&	0.931\\
ZTF18abcysdx&	10.510	$\pm$	0.371&	1.505\\
ZTF18abdbuty&	10.237 $\pm$ 0.348&	0.317\\
ZTF18abddmrf&	9.197 $\pm$		0.946&	3.000\\
ZTF18abdfazk&	9.350 $\pm$		0.271&	1.446\\
    \hline
    \end{tabular}
\end{table}

\begin{table}
    \caption{Table \ref{galaxy_data} continued.}
    \centering
    \begin{tabular}{|l|c|l|}
    \hline
        Object &  Host Galaxy Stellar Mass & DLR\\
        \hline
        ZTF18abdfwur&	9.589 $\pm$		0.376&	0.137\\
        ZTF18abdfydj&	8.801 $\pm$		0.536&	1.140\\
        ZTF18abdkimx&	9.571 $\pm$		0.387&	0.764\\
        ZTF18abdmgab&	10.919 $\pm$		0.325&	17.923\\
        ZTF18abeecwe&	10.118 $\pm$		0.334&	0.918\\
        ZTF18abeegsl&	8.847 $\pm$		0.664&	2.966\\
        ZTF18abetehf&	10.473 $\pm$		0.393&	1.609\\
        ZTF18abetewu&	9.793 $\pm$		0.350&	0.299\\
        ZTF18abfgygp&	9.482 $\pm$		0.504&	0.601\\
        ZTF18abfhaji&	8.623 $\pm$		0.798&	3.552\\
        ZTF18abfhryc&	9.382 $\pm$		0.073&	3.567\\
        ZTF18abfwuwn&	9.250 $\pm$		0.591&	3.794\\
        ZTF18abfzkno&	9.316 $\pm$		0.523&	0.113\\
        ZTF18abgxvra&	10.238 $\pm$		0.410&	7.141\\
        ZTF18abimsyv&	9.353 $\pm$		0.455&	1.792\\
        ZTF18abixjey&	8.281 $\pm$		1.186&	1.301\\
        ZTF18abjstcm&	10.941	$\pm$	0.367&	0.109\\
        ZTF18abjtgdo&	10.555 $\pm$		0.346&	0.823\\
        ZTF18abjtger&	9.798 $\pm$		0.538&	1.110\\
        ZTF18abjvhec&	10.711 $\pm$		0.410&	1.413\\
        ZTF18abkhcrj&	9.993 $\pm$		0.351&	0.183\\
        ZTF18abkhcwl&	9.351 $\pm$		0.338&	1.137\\
        ZTF18abkhdxe&	9.530 $\pm$		0.718&	0.924\\
        ZTF18abkifng&	10.728 $\pm$		0.353&	0.337\\
        ZTF18abkigee&	7.925 $\pm$		1.389&	0.360\\
        ZTF18abklljv&	8.038 $\pm$		1.666&	6.409\\
        ZTF18abkudjo&	10.204 $\pm$		0.390&	1.770\\
        ZTF18abmmkaz&	9.776 $\pm$		0.380&	1.304\\
        ZTF18abmxdhb&	7.819 $\pm$		1.133&	0.702\\
        ZTF18abnvoel&	10.114	$\pm$	0.344&	0.584\\
        ZTF18aboaeqy&	9.549 $\pm$		0.565&	1.043\\
        ZTF18abokpvh&	7.235 $\pm$		1.516&	12.803\\
        ZTF18abpamut&	9.388 $\pm$		0.410&	0.958\\
        ZTF18abpaywm&	9.946 $\pm$		0.429&	1.727\\
        ZTF18abpmmpo&	8.715 $\pm$		0.550&	1.496\\
        ZTF18abptsco&	10.084 $\pm$		0.458&	0.305\\
        ZTF18abpttky&	9.750 $\pm$		0.377&	0.283\\
        ZTF18abqbavl&	10.019 $\pm$		0.448&	2.496\\
        ZTF18abqjvyl&	8.803 $\pm$		0.629&	0.566\\
        ZTF18abrzeym&	9.532 $\pm$		0.366&	0.409\\
        ZTF18abrzrnb&	8.602 $\pm$		1.119&	2.519\\
        ZTF18absdgon&	10.210 $\pm$		0.403&	0.922\\
        ZTF18abslxhz&	9.649 $\pm$		0.575&	0.673\\
        ZTF18abssdpi&	10.355 $\pm$		0.373&	0.162\\
        ZTF18abssuxz&	-&	0.218\\
        ZTF18abtcdfv&	8.277	$\pm$	1.478&	2.419\\
        ZTF18abtnlik&	10.382 $\pm$		0.366&	3.543\\
        ZTF18abtogdl&	8.947 $\pm$		0.483&	0.315\\
        ZTF18abucvbf&	9.200 $\pm$		0.331&	1.151\\
        ZTF18abukmty&	9.326 $\pm$		0.474&	0.433\\
        ZTF18abuqugw&	9.248 $\pm$		0.380&	0.244\\
        ZTF18abvbayb&	10.801 $\pm$		0.377&	3.175\\
        ZTF18abwmuua&	10.527 $\pm$		0.383&	0.254\\
        ZTF18abwnsoc&	8.124 $\pm$		1.481&	13.050\\
        ZTF18abwtops&	10.192	$\pm$	0.419&	4.662\\
        ZTF18abxxssh&	8.645 $\pm$		0.436&	5.987\\
        ZTF18abxygvv&	9.048 $\pm$		0.483&	0.332\\
    \hline
    \end{tabular}
    \label{galaxy_data2}
\end{table}

\begin{table*}
\begin{threeparttable}
    \centering
    \caption{A table summarising the properties of SNe~Ia in our sample. We have included the SNe Ia which could not be well fit by our models, but in these cases the model parameters are not indicative of the true SN parameters.}\label{sn_table1}
    \begin{tabular}{|l|l|l|l|l|l|l|l|l|l|l|}
    \hline
    Name\tnote{a} & z & Model\tnote{b} & Exp. date\tnote{c} &  $\mu$ &  $x_1$ & $c$ & $\chi^2$ & \# unique fits\\
    \hline
ZTF18aailmnv & 0.080     &  EXP\_Ni0.6\_KE2.18\_P3   & 2276211.2 $\pm$ 3.6 & 37.71 $\pm$ 0.4 & 0.4  & -0.06 & 2.3  & 26 \\
ZTF18aaoxryq & 0.0940    &  EXP\_Ni0.5\_KE2.18\_P3   & 2247096.7 $\pm$ 2.7 & 37.95 $\pm$ 0.3 & -0.2 & 0.02  & 1.1  & 10  \\
ZTF18aapqwyv & 0.0560    &  EXP\_Ni0.4\_KE0.78\_P21  & 2327864.7 $\pm$ 5.0 & 36.85 $\pm$ 0.5 & -1.7 & 0.17  & 8.0  & 40 \\
ZTF18aapsedq &  0.0650   &  EXP\_Ni0.4\_KE1.10\_P4.4 & 2308194.3 $\pm$ 1.1 & 37.16 $\pm$ 0.3 & -0.1 & -0.09 & 0.6  & 4  \\
ZTF18aaqcozd & 0.0732    &  DPL\_Ni0.4\_KE2.18\_P3   & 2290994.3 $\pm$ 3.0 & 37.46 $\pm$ 0.4 & -1.2 & -0.1  & 6.3  & 19 \\
ZTF18aaqcqkv & 0.1174     &  DPL\_Ni0.4\_KE2.18\_P3   & 2194847.7 $\pm$ 4.2 & 38.64 $\pm$ 0.6 & -1.3 & -0.04 & 2.2  & 14 \\
ZTF18aaqcqvr & 0.070     &  DPL\_Ni0.4\_KE2.18\_P4.4 & 2297418.5 $\pm$ 2.1 & 37.24 $\pm$ 0.3 & 0.2  & -0.02 & 2.8  & 16 \\
ZTF18aaqcugm & 0.066     &  DPL\_Ni0.4\_KE2.18\_P4.4 & 2319092.1 $\pm$ 2.0 & 36.95 $\pm$ 0.3 & -1.1 & -0.09 & 7.0  & 5  \\
ZTF18aaqffyp & 0.070     &  DPL\_Ni0.4\_KE1.81\_P3   & 2297416.1 $\pm$ 4.0 & 37.36 $\pm$ 0.4 & 0.9  & 0.06  & 3.9  & 25 \\
ZTF18aaqnrum & 0.066    &  DPL\_Ni0.4\_KE2.18\_P4.4 & 2306038.3 $\pm$ 2.5 & 37.35 $\pm$ 0.3 & 0.1  & -0.03 & 3.3  & 6  \\
\textbf{ZTF18aaqqoqs} & 0.082    &  EXP\_Ni0.5\_KE1.10\_P4.4 & 2271941.5 $\pm$ 1.2 & 37.46 $\pm$ 0.0 & 1.2  & -0.01 & 0.9  & 2  \\
ZTF18aarldnh & 0.1077  &  EXP\_Ni0.4\_KE2.18\_P3   & 2219188.3 $\pm$ 1.7 & 38.22 $\pm$ 0.4 & -1.0 & 0.07  & 0.7  & 8  \\
ZTF18aarqnje & 0.117    &  EXP\_Ni0.6\_KE2.18\_P3   & 2200752.6 $\pm$ 1.8 & 38.29 $\pm$ 0.0 & 0.2  & -0.18 & 1.0  & 4  \\
ZTF18aasdted & 0.0182  &  EXP\_Ni0.4\_KE1.40\_P3   & 2414424.0 $\pm$ 1.3 & 34.64 $\pm$ 0.2 & 0.8  & 0.16  & 13.9 & 7  \\
ZTF18aaslhxt & 0.0550     &  EXP\_Ni0.6\_KE2.18\_P4.4 & 2319098.2 $\pm$ 1.4 & 36.98 $\pm$ 0.1 & 0.3  & -0.11 & 1.7  & 3  \\
ZTF18aatzygk & 0.077    &  EXP\_Ni0.5\_KE2.18\_P3   & 2282492.9 $\pm$ 4.6 & 37.75 $\pm$ 0.5 & 0.0  & 0.05  & 2.3  & 46 \\
ZTF18aauhxce & 0.0831  &  EXP\_Ni0.4\_KE1.10\_P3   & 2269684.1 $\pm$ 2.3 & 37.54 $\pm$ 0.4 & 1.3  & 0.12  & 1.8  & 9  \\
ZTF18aaumeys & 0.0365  &  EXP\_Ni0.5\_KE1.10\_P3   & 2369304.1 $\pm$ 2.0 & 35.85 $\pm$ 0.4 & 0.7  & -0.02 & 3.6  & 15 \\
ZTF18aaumlfl & 0.0874  &  DPL\_Ni0.4\_KE2.18\_P4.4 & 2260727.9 $\pm$ 2.0 & 37.87 $\pm$ 0.3 & -1.1 & -0.09 & 1.5  & 9  \\
ZTF18aaunfqq & 0.0711  &  EXP\_Ni0.5\_KE2.18\_P3   & 2295092.5 $\pm$ 1.8 & 37.65 $\pm$ 0.4 & -1.2 & 0.01  & 1.3  & 10 \\
ZTF18aauocnw & 0.102    &  EXP\_Ni0.6\_KE2.18\_P3   & 2230717.4 $\pm$ 3.8 & 37.97 $\pm$ 0.2 & 0.1  & -0.05 & 2.7  & 49 \\
ZTF18aavrwhu & 0.0620  &  EXP\_Ni0.5\_KE0.78\_P4.4 & 2314718.7 $\pm$ 1.6 & 36.83 $\pm$ 0.1 & 1.2  & -0.02 & 2.2  & 3  \\
ZTF18aavrzxp & 0.069    &  DPL\_Ni0.4\_KE2.18\_P4.4 & 2299582.2 $\pm$ 2.3 & 37.45 $\pm$ 0.3 & -1.4 & -0.07 & 2.7  & 7  \\
ZTF18aawjywv & 0.0509     &  DPL\_Ni0.4\_KE2.18\_P3   & 2363707.4 $\pm$ 3.8 & 36.68 $\pm$ 0.6 & -1.5 & -0.01 & 2.4  & 23 \\
ZTF18aawurud & 0.0531     &  EXP\_Ni0.4\_KE2.18\_P3   & 2341246.2 $\pm$ 3.0 & 36.64 $\pm$ 0.6 & 0.3  & 0.14  & 1.2  & 16 \\
ZTF18aaxakhh & 0.117    &  EXP\_Ni0.5\_KE2.18\_P3   & 2200759.5 $\pm$ 2.5 & 38.33 $\pm$ 0.2 & 0.5  & -0.08 & 1.3  & 9  \\
ZTF18aaxcntm & 0.0269 &  EXP\_Ni0.4\_KE1.68\_P100 & 2394119.0 $\pm$ 4.2 & 35.51 $\pm$ 0.4 & -1.5 & 0.1   & 27.0 & 26 \\
ZTF18aaxdrjn & 0.0340   &  EXP\_Ni0.5\_KE0.50\_P9.7 & 2377657.9 $\pm$ 3.3 & 35.56 $\pm$ 0.6 & -1.9 & -0.08 & 4.1  & 32 \\
ZTF18aaxqyki & 0.1003      &  EXP\_Ni0.5\_KE1.40\_P3   & 2234787.1 $\pm$ 2.2 & 37.97 $\pm$ 0.2 & 0.9  & -0.05 & 1.2  & 10  \\
ZTF18aaxrvzj & 0.114    &  EXP\_Ni0.5\_KE1.81\_P3   & 2206700.8 $\pm$ 1.8 & 38.23 $\pm$ 0.0 & 1.4  & -0.01 & 0.9  & 6  \\
ZTF18aaxsioa & 0.0315     &  EXP\_Ni0.4\_KE2.18\_P3   & 2386669.7 $\pm$ 0.2 & 35.69 $\pm$ 0.0 & -1.5 & 0.05  & 2.3  & 1  \\
ZTF18aaxvpsw & 0.0916   &  EXP\_Ni0.5\_KE2.18\_P4.4 & 2251985.8 $\pm$ 1.7 & 37.85 $\pm$ 0.1 & 0.0  & -0.01 & 0.6  & 6  \\
ZTF18aaxwjmp & 0.084    &  DPL\_Ni0.4\_KE1.81\_P4.4 & 2267775.6 $\pm$ 1.5 & 37.56 $\pm$ 0.3 & 0.4  & -0.01 & 0.9  & 11  \\
ZTF18aaydmkh & 0.077    &  EXP\_Ni0.4\_KE2.18\_P3   & 2282515.7 $\pm$ 1.8 & 37.83 $\pm$ 0.1 & 0.3  & 0.19  & 2.8  & 4  \\
\textbf{ZTF18aayjvve} & 0.0474  &  EXP\_Ni0.4\_KE1.68\_P3   & 2347316.1 $\pm$ 0.4 & 36.43 $\pm$ 0.1 & -0.1 & 0.06  & 2.1  & 4  \\
ZTF18aaytovs & 0.0746     &  EXP\_Ni0.5\_KE1.81\_P3   & 2276179.5 $\pm$ 2.8 & 37.25 $\pm$ 0.1 & 2.0  & 0.12  & 1.4  & 10 \\
ZTF18aazabmh & 0.0746    &  EXP\_Ni0.4\_KE2.18\_P3   & 2291027.2 $\pm$ 1.0 & 37.67 $\pm$ 0.1 & -0.3 & 0.11  & 1.4  & 3  \\
ZTF18aazblzy & 0.0653     &  DPL\_Ni0.4\_KE2.18\_P4.4 & 2319127.3 $\pm$ 0.6 & 37.29 $\pm$ 0.0 & -1.7 & -0.07 & 1.7  & 1  \\
ZTF18aazcoob & 0.0845    &  EXP\_Ni0.5\_KE1.10\_P3   & 2259443.2 $\pm$ 2.2 & 37.62 $\pm$ 0.1 & 0.7  & 0.06  & 2.2  & 5  \\
ZTF18aazixbw & 0.0594   &  EXP\_Ni0.4\_KE2.18\_P3   & 2320440.9 $\pm$ 0.5 & 37.16 $\pm$ 0.0 & -1.6 & 0.05  & 1.3  & 1  \\
ZTF18aazjztm & 0.0721    &  EXP\_Ni0.5\_KE2.18\_P3   & 2293166.6 $\pm$ 1.3 & 37.77 $\pm$ 0.3 & -1.7 & 0.09  & 1.3  & 4  \\
ZTF18aazsabq & 0.060     &  EXP\_Ni0.4\_KE2.18\_P3   & 2319129.2 $\pm$ 1.2 & 37.01 $\pm$ 0.3 & -1.2 & 0.03  & 1.5  & 3  \\
ZTF18abatffv & 0.143    &  EXP\_Ni0.5\_KE1.40\_P4.4 & 2150720.4 $\pm$ 4.4 & 38.85 $\pm$ 0.3 & 1.0  & -0.01 & 1.2  & 46 \\
ZTF18abauprj & 0.0242    &  EXP\_Ni0.6\_KE2.18\_P4.4 & 2391325.3 $\pm$ 1.4 & 34.89 $\pm$ 0.2 & 1.3  & -0.01 & 6.9  & 4  \\
ZTF18abaxlpi & 0.0642   &  EXP\_Ni0.5\_KE2.18\_P3   & 2309111.6 $\pm$ 0.7 & 37.46 $\pm$ 0.3 & 0.1  & 0.06  & 0.7  & 3  \\
ZTF18abbpeqo & 0.0667  &  EXP\_Ni0.5\_KE2.18\_P4.4 & 2304394.4 $\pm$ 1.8 & 37.29 $\pm$ 0.3 & -0.1 & 0.02  & 0.8  & 7  \\
ZTF18abbvsiv & 0.051    &  DPL\_Ni0.4\_KE2.18\_P4.4 & 2338991.4 $\pm$ 1.9 & 36.6  $\pm$ 0.3 & -1.7 & 0.01  & 2.3  & 17 \\
ZTF18abcecfi & 0.079    &  EXP\_Ni0.4\_KE2.18\_P3   & 2278297.1 $\pm$ 1.4 & 37.72 $\pm$ 0.3 & 0.1  & 0.09  & 1.1  & 6  \\
ZTF18abcflnz & 0.0273    &  EXP\_Ni0.5\_KE0.78\_P4.4 & 2400670.4 $\pm$ 0.5 & 35.07 $\pm$ 0.3 & 0.1  & -0.03 & 5.8  & 4  \\
ZTF18abckujg & 0.075    &  EXP\_Ni0.5\_KE1.40\_P3   & 2286775.8 $\pm$ 2.1 & 37.52 $\pm$ 0.3 & 0.5  & 0.03  & 1.1  & 14 \\
ZTF18abckujq & 0.0638  &  EXP\_Ni0.5\_KE1.40\_P9.7 & 2310786.4 $\pm$ 2.6 & 36.89 $\pm$ 0.3 & 0.5  & -0.02 & 2.5  & 18 \\
ZTF18abclalx & 0.105    &  DPL\_Ni0.4\_KE2.18\_P4.4 & 2224690.0 $\pm$ 2.0 & 38.03 $\pm$ 0.3 & -1.4 & -0.08 & 1.7  & 6  \\
ZTF18abcrxoj & 0.0309   &  EXP\_Ni0.5\_KE2.18\_P4.4 & 2384604.8 $\pm$ 1.2 & 35.69 $\pm$ 0.3 & -1.3 & 0.04  & 1.7  & 3  \\
ZTF18abcsgvj & 0.060     &  DPL\_Ni0.4\_KE2.18\_P3   & 2319138.9 $\pm$ 1.7 & 37.31 $\pm$ 0.2 & -1.7 & 0.04  & 1.4  & 4  \\
ZTF18abcysdx & 0.066    &  EXP\_Ni0.4\_KE0.78\_P21  & 2306083.8 $\pm$ 5.3 & 37.22 $\pm$ 0.5 & 0.6  & 0.16  & 3.7  & 29 \\
ZTF18abdbuty & 0.059    &  EXP\_Ni0.5\_KE2.18\_P4.4 & 2321331.2 $\pm$ 2.3 & 37.18 $\pm$ 0.4 & -0.8 & 0.01  & 1.5  & 16 \\
ZTF18abddmrf & 0.163    &  EXP\_Ni0.6\_KE2.18\_P4.4 & 2113743.2 $\pm$ 5.0 & 39.07 $\pm$ 0.3 & 2.5  & 0.08  & 2.7  & 50\\
    \hline
    \end{tabular}
    \begin{tablenotes}
    \footnotesize 
        \item[a] The names of SNe Ia with a detected flux excess are highlighted in bold.
        \item[b] Filename of model, encapsulating all the model parameters (\nick\ mass, $P$-value, kinetic energy, and density profile).
        \item[c] Estimated explosion date from fit in the rest frame of the SN.
      \item[d] $x_1$ values, $c$ colours, and redshifts ($z$) are taken from \cite{Yao2019}. $z$ is noted to 4 decimal places if the value is taken from NED, or measured from the galaxy spectrum obtained by \cite{Yao2019}, or measured from a SN spectrum where the host H$\alpha$ line can be identified. If $z$ is taken from a SNID fit it is noted to 3 decimal places.
    \end{tablenotes}
    \end{threeparttable}
\end{table*}

\begin{table*}
\begin{threeparttable}
    \centering
    \caption{Continuation of Table \ref{sn_table1}}\label{sn_table2}
    \begin{tabular}{|l|l|l|l|l|l|l|l|r|l|l|}
    \hline
    Name & z & Model & Exp. date &  $\mu$ &  $x_1$ & $c$ & $\chi^2$ & \# unique fits\\
\hline
ZTF18abdefet & 0.074    &  EXP\_Ni0.4\_KE2.18\_P3   & 2288909.8 $\pm$ 2.3 & 37.7  $\pm$ 0.1 & -0.1 & 0.14  & 3.5 & 6  \\
\textbf{ZTF18abdfazk} & 0.084    &  DPL\_Ni0.4\_KE2.18\_P4.4 & 2267794.3 $\pm$ 1.8 & 37.56 $\pm$ 0.4 & -0.3 & -0.06 & 1.1 & 10  \\
\textbf{ZTF18abdfwur} & 0.070     &  DPL\_Ni0.4\_KE1.81\_P3   & 2297469.8 $\pm$ 3.4 & 37.36 $\pm$ 0.4 & -0.5 & 0.04  & 1.3 & 24 \\
ZTF18abdfydj & 0.076    &  DPL\_Ni0.4\_KE2.18\_P4.4 & 2284658.7 $\pm$ 0.8 & 37.29 $\pm$ 0.0 & 0.2  & -0.04 & 1.8 & 1  \\
ZTF18abdkimx & 0.077    &  DPL\_Ni0.4\_KE2.18\_P4.4 & 2282540.1 $\pm$ 2.1 & 37.49 $\pm$ 0.4 & 0.0  & -0.04 & 1.6 & 10  \\
ZTF18abdmgab & 0.0803     &  EXP\_Ni0.4\_KE1.68\_P100 & 2276198.2 $\pm$ 5.0 & 37.97 $\pm$ 0.4 & -2.3 & 0.12  & 4.6 & 28 \\
ZTF18abeecwe & 0.0393     &  EXP\_Ni0.4\_KE1.81\_P3   & 2363751.7 $\pm$ 1.6 & 35.84 $\pm$ 0.4 & -0.6 & 0.02  & 1.8 & 11 \\
ZTF18abeegsl & 0.072    &  EXP\_Ni0.4\_KE1.68\_P100 & 2293187.5 $\pm$ 4.0 & 37.64 $\pm$ 0.4 & -2.2 & 0.16  & 9.9 & 19 \\
ZTF18abetehf & 0.0649 &  DPL\_Ni0.4\_KE2.18\_P3   & 2308423.1 $\pm$ 3.0 & 36.97 $\pm$ 0.3 & -1.4 & -0.14 & 2.2 & 15 \\
ZTF18abetewu & 0.077    &  EXP\_Ni0.5\_KE2.18\_P3   & 2282544.3 $\pm$ 2.0 & 37.7 $\pm$ 0.5 & 0.3  & 0.11  & 1.0 & 10 \\
ZTF18abfgygp & 0.064    &  DPL\_Ni0.4\_KE1.81\_P4.4 & 2310434.7 $\pm$ 1.9 & 36.94 $\pm$ 0.4 & 0.1  & -0.07 & 1.0 & 6  \\
ZTF18abfhaji & 0.084    &  EXP\_Ni0.5\_KE1.68\_P4.4 & 2267805.7 $\pm$ 2.0 & 37.82 $\pm$ 0.3 & -0.2 & -0.04 & 0.8 & 14 \\
ZTF18abfhryc & 0.0323    &  EXP\_Ni0.4\_KE1.81\_P4.4 & 2386703.0 $\pm$ 0.6 & 35.36 $\pm$ 0.3 & 0.5  & 0.01  & 4.7 & 6  \\
ZTF18abfwuwn & 0.109    &  EXP\_Ni0.6\_KE2.18\_P3   & 2216686.0 $\pm$ 1.6 & 38.12 $\pm$ 0.0 & 0.6  & -0.03 & 1.9 & 6  \\
ZTF18abfzkno & 0.100      &  EXP\_Ni0.5\_KE2.18\_P4.4 & 2234821.0 $\pm$ 2.9 & 38.09 $\pm$ 0.3 & -0.6 & -0.07 & 0.8 & 16 \\
ZTF18abgxvra & 0.104    &  EXP\_Ni0.6\_KE2.18\_P4.4 & 2226728.3 $\pm$ 2.4 & 38.01 $\pm$ 0.0 & 0.8  & -0.04 & 4.1 & 8  \\
ZTF18abimsyv & 0.088    &  EXP\_Ni0.5\_KE1.40\_P4.4 & 2259479.0 $\pm$ 1.6 & 37.63 $\pm$ 0.1 & 1.0  & -0.03 & 0.9 & 6  \\
ZTF18abixjey & 0.1218     &  EXP\_Ni0.6\_KE2.18\_P3   & 2194920.6 $\pm$ 2.8 & 38.47 $\pm$ 0.1 & 0.8  & -0.07 & 1.3 & 12 \\
ZTF18abjtgdo & 0.0741  &  EXP\_Ni0.5\_KE1.40\_P3   & 2289086.7 $\pm$ 3.6 & 37.44 $\pm$ 0.4 & -0.8 & 0.0   & 0.6 & 27 \\
ZTF18abjtger & 0.107    &  EXP\_Ni0.5\_KE2.18\_P4.4 & 2220701.0 $\pm$ 3.4 & 38.08 $\pm$ 0.2 & 0.8  & -0.01 & 2.6 & 17 \\
ZTF18abjvhec & 0.0570   &  EXP\_Ni0.5\_KE1.40\_P4.4 & 2325972.6 $\pm$ 2.0 & 36.63 $\pm$ 0.2 & 0.4  & -0.04 & 1.3 & 5  \\
ZTF18abkhcrj & 0.0383  &  EXP\_Ni0.4\_KE2.18\_P3   & 2367803.3 $\pm$ 1.5 & 36.34 $\pm$ 0.1 & 0.9  & 0.24  & 6.7 & 4  \\
ZTF18abkhcwl & 0.0317  &  EXP\_Ni0.5\_KE1.10\_P4.4 & 2382929.2 $\pm$ 2.2 & 35.49 $\pm$ 0.4 & 0.1  & -0.08 & 5.7 & 12 \\
ZTF18abkhdxe & 0.104    &  EXP\_Ni0.4\_KE2.18\_P3   & 2226737.4 $\pm$ 2.6 & 38.23 $\pm$ 0.5 & 0.7  & 0.12  & 1.0 & 10 \\
ZTF18abkifng & 0.0880    &  EXP\_Ni0.5\_KE2.18\_P4.4 & 2259489.0 $\pm$ 3.3 & 37.8 $\pm$ 0.4 & 0.3  & -0.04 & 0.9 & 23 \\
ZTF18abkigee & 0.0936      &  EXP\_Ni0.5\_KE1.68\_P3   & 2234838.4 $\pm$ 2.9 & 38.15 $\pm$ 0.5 & 1.0  & -0.03 & 1.1 & 20 \\
ZTF18abklljv & 0.141    &  EXP\_Ni0.6\_KE2.18\_P3   & 2154530.4 $\pm$ 1.9 & 38.73 $\pm$ 0.0 & 1.2  & -0.1  & 0.5 & 3  \\
ZTF18abkudjo & 0.0921  &  EXP\_Ni0.4\_KE1.81\_P4.4 & 2250882.8 $\pm$ 3.3 & 37.73 $\pm$ 0.4 & 0.9  & 0.02  & 0.9 & 24 \\
ZTF18abmmkaz & 0.063    &  EXP\_Ni0.5\_KE1.40\_P4.4 & 2312638.7 $\pm$ 1.9 & 36.95 $\pm$ 0.3 & 0.7  & -0.07 & 1.8 & 7  \\
ZTF18abmxdhb & 0.070     &  DPL\_Ni0.4\_KE1.81\_P4.4 & 2297515.6 $\pm$ 1.9 & 37.31 $\pm$ 0.5 & 1.3  & 0.01  & 1.0 & 16 \\
ZTF18abnvoel & 0.083    &  EXP\_Ni0.6\_KE2.18\_P3   & 2269937.5 $\pm$ 1.7 & 37.49 $\pm$ 0.0 & 0.6  & -0.01 & 4.6 & 4  \\
ZTF18aboaeqy & 0.129    &  EXP\_Ni0.5\_KE2.18\_P4.4 & 2177447.1 $\pm$ 3.4 & 38.52 $\pm$ 0.1 & 0.3  & -0.03 & 1.4 & 10  \\
ZTF18abokpvh & 0.081    &  EXP\_Ni0.4\_KE2.18\_P4.4 & 2274140.1 $\pm$ 1.8 & 37.43 $\pm$ 0.4 & 0.8  & -0.05 & 1.1 & 9  \\
\textbf{ZTF18abpamut} & 0.064    &  EXP\_Ni0.4\_KE1.10\_P3   & 2310477.5 $\pm$ 3.3 & 37.28 $\pm$ 0.5 & 0.8  & 0.1   & 1.8 & 7  \\
ZTF18abpaywm & 0.040     &  EXP\_Ni0.4\_KE1.81\_P3   & 2363797.4 $\pm$ 4.3 & 36.44 $\pm$ 0.5 & 0.6  & 0.23  & 6.9 & 9  \\
ZTF18abpmmpo & 0.076    &  EXP\_Ni0.5\_KE1.40\_P4.4 & 2284709.7 $\pm$ 1.2 & 37.29 $\pm$ 0.0 & 1.5  & -0.01 & 2.1 & 3  \\
ZTF18abptsco\tnote{*} & 0.12     &  DPL\_Ni0.4\_KE2.18\_P3   & 2194951.2 $\pm$ 4.2 & 38.6  $\pm$ 0.4 & -0.4 & 0.07  & 1.4 & 15 \\
ZTF18abpttky & 0.084    &  DPL\_Ni0.4\_KE2.18\_P3   & 2267849.2 $\pm$ 2.8 & 37.82 $\pm$ 0.3 & -1.3 & 0.02  & 1.7 & 15 \\
ZTF18abqbavl & 0.1392     &  EXP\_Ni0.5\_KE1.40\_P3   & 2156439.7 $\pm$ 3.6 & 38.7 $\pm$ 0.2 & 0.7  & -0.03 & 0.9 & 18 \\
ZTF18abqjvyl & 0.0835   &  DPL\_Ni0.4\_KE1.68\_P4.4 & 2260133.4 $\pm$ 3.1 & 37.59 $\pm$ 0.5 & 0.3  & -0.06 & 0.6 & 17 \\
ZTF18abrzeym & 0.055    &  EXP\_Ni0.4\_KE0.78\_P21  & 2330192.7 $\pm$ 4.9 & 36.81 $\pm$ 0.5 & 0.4  & 0.23  & 9.9 & 37 \\
ZTF18abrzrnb & 0.120     &  EXP\_Ni0.5\_KE2.18\_P3   & 2194959.4 $\pm$ 3.3 & 38.35 $\pm$ 0.1 & 0.4  & -0.02 & 1.4 & 9  \\
ZTF18absdgon & 0.0620     &  EXP\_Ni0.4\_KE2.18\_P3   & 2255374.0 $\pm$ 1.4 & 37.17 $\pm$ 0.3 & -0.3 & 0.17  & 5.0 & 3  \\
ZTF18abslxhz & 0.134    &  EXP\_Ni0.6\_KE2.18\_P3   & 2167861.9 $\pm$ 1.2 & 38.61 $\pm$ 0.0 & 0.7  & -0.08 & 1.3 & 2  \\
ZTF18abssdpi & 0.103    &  EXP\_Ni0.4\_KE2.18\_P3   & 2228791.2 $\pm$ 2.9 & 38.25 $\pm$ 0.4 & -0.3 & 0.2   & 2.0 & 9  \\
ZTF18abssuxz & 0.0649   &  EXP\_Ni0.5\_KE2.18\_P4.4 & 2308536.0 $\pm$ 2.0 & 37.23 $\pm$ 0.3 & -1.1 & 0.02  & 0.7 & 9  \\
ZTF18abtcdfv & 0.140     &  EXP\_Ni0.6\_KE2.18\_P3   & 2156453.8 $\pm$ 1.1 & 38.71 $\pm$ 0.0 & 1.1  & -0.05 & 1.2 & 1  \\
ZTF18abtnlik & 0.084    &  EXP\_Ni0.4\_KE2.18\_P3   & 2267861.3 $\pm$ 0.9 & 37.6 $\pm$ 0.3 & -1.2 & 0.06  & 0.9 & 3  \\
ZTF18abtogdl & 0.100      &  EXP\_Ni0.5\_KE1.40\_P4.4 & 2234873.4 $\pm$ 2.5 & 37.92 $\pm$ 0.2 & 0.3  & -0.06 & 1.1 & 10 \\
ZTF18abucvbf & 0.0549     &  EXP\_Ni0.6\_KE2.18\_P4.4 & 2341301.2 $\pm$ 1.5 & 36.85 $\pm$ 0.3 & 0.0  & -0.07 & 4.3 & 9  \\
ZTF18abukmty & 0.104    &  EXP\_Ni0.5\_KE1.81\_P3   & 2226778.4 $\pm$ 2.0 & 38.14 $\pm$ 0.2 & 0.5  & 0.0   & 1.3 & 8  \\
ZTF18abuqugw & 0.0313  &  DPL\_Ni0.4\_KE2.18\_P4.4 & 2384357.8 $\pm$ 2.4 & 35.38 $\pm$ 0.3 & -1.3 & -0.12 & 4.7 & 11  \\
ZTF18abvbayb & 0.132    &  EXP\_Ni0.5\_KE2.18\_P4.4 & 2171699.8 $\pm$ 1.8 & 38.57 $\pm$ 0.0 & -0.2 & -0.05 & 0.5 & 5  \\
ZTF18abwmuua & 0.0643  &  EXP\_Ni0.4\_KE2.18\_P4.4 & 2309976.4 $\pm$ 4.8 & 37.08 $\pm$ 0.6 & -1.2 & -0.01 & 0.5 & 81 \\
ZTF18abwnsoc & 0.099    &  EXP\_Ni0.5\_KE1.68\_P3   & 2236916.6 $\pm$ 1.8 & 38.07 $\pm$ 0.3 & 0.3  & 0.01  & 0.4 & 10 \\
ZTF18abwtops & 0.101    &  EXP\_Ni0.5\_KE2.18\_P4.4 & 2232851.4 $\pm$ 2.5 & 38.37 $\pm$ 0.5 & -1.4 & -0.05 & 0.8 & 19 \\
\textbf{ZTF18abxxssh} & 0.064    &  EXP\_Ni0.6\_KE1.81\_P4.4 & 2310503.7 $\pm$ 3.7 & 37.57 $\pm$ 0.4 & 1.5  & -0.02 & 3.2 & 44 \\
ZTF18abxygvv & 0.079    &  EXP\_Ni0.6\_KE2.18\_P3   & 2278384.0 $\pm$ 1.6 & 37.55 $\pm$ 0.3 & -0.1 & -0.04 & 0.6 & 7 \\
    \hline
    \end{tabular}
    \begin{tablenotes}
    \footnotesize 
        \item[*] The redshift of ZTF18abptsco was reported by ATel 12052 \citep{Gomez2018} and was shown with two decimal places.
    \end{tablenotes}
    \end{threeparttable}
\end{table*}

\bsp	
\label{lastpage}
\end{document}